\documentclass[12pt]{article}
\usepackage{latexsym}
\usepackage{pictex}
\usepackage{amssymb}
\usepackage{graphicx}
\makeatletter
\@addtoreset{equation}{section}
\makeatother

\newtheorem{theorem}{Theorem}
\newtheorem{proposition}{Proposition}
\newcommand{\simlt}
{\mbox{\raisebox{-0.5ex}{$\textstyle \; \sim$}
\raisebox{ 0.8ex}{$\textstyle  \!\!\!\!\!\!\! <$  }}}
\newcommand{\simgt}
{\mbox{\raisebox{-0.5ex}{$\textstyle \; \sim$}
\raisebox{ 0.8ex}{$\textstyle  \!\!\!\!\!\!\! >$  }}}
\begin{document}
\title{
 Black-Hole Phenomenology
 }
\author{
Neven Bili\'c
 \\
 Rudjer Bo\v{s}kovi\'{c} Institute, 10002 Zagreb, Croatia \\
E-mail: bilic@thphys.irb.hr
}
\maketitle
%
\begin{abstract}
 
This set of lectures is an introduction to 
black-hole astrophysics. The emphasis is made on
the phenomenology of X-ray binaries and of
supermassive compact objects at galactic centers.
\end{abstract}
%

\tableofcontents
\section{Introduction}
\label{introduction}
Black holes (BH) are the most fascinating objects predicted by 
general relativity. 
There exist about 20 confirmed candidates
\cite{rem} for astrophysical BHs in
the mass range 5 - 20 $M_\odot$ and about three dozen 
supermassive BH
candidates \cite{fer1} in the mass range $10^6$ - $10^{9.5}$ $M_\odot$.

 Unfortunately, there exists as yet no direct evidence  for astrophysical BHs
 \footnote{See A.~M\"uller's lecture at this School}.
At present we only hope that the black-hole
paradigm may  be proved
or ruled out by comparing BH candidates with
credible alternatives. 
 Fortunately, BHs are dark and compact, which narrows the list
 of possible alternatives among standard astrophysical objects. 
 For a stellar-mass BH, the only standard astrophysical alternative is
 {\em neutron stars} (NS). It is therefore quite important to
 understand the properties of NSs and their observational 
 distinction to BHs. Therefore,  considerable attention 
will be devoted here to NSs.
 
These lecture notes
are in large part based on the standard text books 
Misner, Thorne, and Wheeler \cite{mis},
 Shapiro and Teukolsky \cite{sha},  
 Wald \cite{wal}, and
 Carroll and Ostlie \cite{carro},
 and on the  review articles by
 Townsend \cite{tau} and
 Narayan \cite{nar}.
 The exception is section \ref{alternatives}
 which is based on original articles.
Efforts are made to provide citations to original papers  wherever appropriate
but the list of references is by no means complete. A number of interesting talks
may be found at the site
  of a recent conference on supermassive black holes \cite{santa}.
 
We use the positive-time negative-space signature convention, i.e.,
 (+,- - -) and
we  mostly use the so-called natural units in which
$c=\hbar = G=1$. In these units, the physical quantities are expressed
in powers of the Planck mass $m_{\rm Pl}=\sqrt{\hbar c/G}$, Planck length
$l_{\rm Pl}=\sqrt{\hbar G/c^3}$, or Planck time $t_{\rm Pl}=l_{\rm Pl}/c$.

 
\section{Preliminaries}

\subsection{Spherical Configurations}
Consider the space time metric
\begin{equation}
ds^2=g_{\mu\nu}dx^\mu dx^\nu,
\label{eq0}
\end{equation}
in which the metric tensor $g_{\mu\nu}$
is time independent.
In general relativity we distingush between static and stationary
metric.
Stationary configurations are described by the metric 
coefficients that do not depend on time.
For static configurations we also require the time reversal
 ($t\rightarrow -t$) invariance of the metric.
In this case, the mixed components $g_{0i}$ must vanish.

Static fluid configurations  are spherical.
The most general metric generated by a spherical mass distribution
is  of the form
\begin{equation}
ds^2=\xi^2 dt^2 -\lambda^2 dr^2 -
     r^2(d\theta^2+\sin^2 \theta d\phi^2),
\label{eq00}
\end{equation}
where $\xi$ and $\lambda$ are functions of $r$ only.
The function $\xi$, called the  ``lapse function", may be represented in terms of the
gravitational potential 
\begin{equation}
\xi=e^{\varphi (r)},
\label{eq01}
\end{equation}
and the function $\lambda$ is related to the enclosed mass ${\cal M}(r)$
\begin{equation}
\lambda=\left(1-\frac{2{\cal M}(r)}{r}\right)^{-1/2} .
\label{eq10}
\end{equation}
It may be easily shown \cite{tol} that Einstein's equations 
reduce to two nontrivial independent 
equations 
\begin{equation}
\frac{2}{\xi}\frac{d\xi}{dr}=-8\pi r \lambda^2 {T^r}_r+
\frac{\lambda^2-1}{r} ,
\label{eq1}
\end{equation} 
\begin{equation}
\frac{2}{\lambda}\frac{d\lambda}{dr}=8\pi r \lambda^2 {T^0}_0-
\frac{\lambda^2-1}{r} .
\label{eq2}
\end{equation} 
For a perfect fluid
\begin{equation}
{T^r}_r=-p  \, ,   \;\;\;\;\;\;\;\ {T^0}_0=\rho.
\label{eq3}
\end{equation} 
Using (\ref{eq10}),
Einstein's field equations take the form
\begin{equation}
\frac{d\xi}{dr}=\xi\frac{{\cal M}+4\pi r^3 p}{r(r-2{\cal{M}})} \, ,
\label{eq40}
\end{equation}
\begin{equation}
\frac{d{\cal{M}}}{dr}=4\pi r^2 \rho .
\label{eq42}
\end{equation}
The latter
may be written in the form
\begin{equation}
{\cal M}(r)=\int^r_0 dr'\, 4\pi r'^2  \rho(r')   \, ,
\label{eq11}
\end{equation}
 which shows that the function ${\cal M}$ may
 indeed be interpreted as an enclosed mass.
 
 Finally, if we impose the particle number conservation constraint (\ref{eq26}),
 we have
\begin{equation}
\int_0^Rdr\, 4\pi r^2 (1-2{\cal{M}}/r)^{-1/2}\, n(r)=N ,
\label{eq45}
\end{equation}
where we have employed the spherical symmetry to
replace the proper volume integral as
\begin{equation}
\int_{\Sigma} u^{\mu}d\Sigma_{\mu}
= \int_0^R dr 4\pi r^2 \lambda .
\label{eq80}
\end{equation}
 
\subsection{Schwarzschild Solution}
\label{sch}
Assume the absence of matter in the region  $r>R$.
In this region
equations (\ref{eq40}) and (\ref{eq42}) may be easily
solved. One finds 
\begin{equation}
{\cal M}(r)=M=\mbox{const.}  \, ,
\label{eq12}
\end{equation} 
\begin{equation}
\xi(r)=\left(1-\frac{2M}{r}\right)^{1/2}.
\label{eq46}
\end{equation} 
 The  metric takes the form
\begin{equation}
ds^2=\left(1-\frac{2M}{r}\right)dt^2 - \left(1-\frac{2M}{r}\right)^{-1}
dr^2 - r^2(d\theta^2+\sin^2 \theta d\phi^2).
\label{eq47}
\end{equation} 
 This is known as the {\em Schwarzschild metric}. 
 The constant $M$ is the mass of the source. 
 In the weak field limit $r\gg M$, we obtain the Newtonian potential
 \begin{equation}
\varphi=\ln \xi \approx -\frac{M}{r}\, .
 \end{equation} 
  This metric describes the gravitational field outside of any
 spherical object of mass $M$, including a black hole.
 The sphere with radius  $r=2M$, at which the metric diverges,
 is the BH {\em event horizon}. 
 \subsubsection{Birkhoff's Theorem}
 \label{bir}
 \begin{theorem}
  The exterior spacetime of {\bf all} spherical gravitating bodies
 (not necessarily sta\-tic) is described by the Schwarzschild metric.
 \end{theorem}
 The proof is simple. See, e.g., Misner et al. \cite{mis}.
\subsection{Spherical Stars}
Generally, we assume that $\rho$ and $p$ satisfy an equation of state,
e.g., in the form given by (\ref{eq97}) and (\ref{eq98}),
and are generally functions of local $T$ and $\mu$
which in turn depend on $\xi$ through the Tolman equations (\ref{eq21}).
Numerical integration of (\ref{eq40}) and (\ref{eq42})
from $r=0$ to some radius $R$ 
 with the boundary conditions
\begin{equation}
\xi(R)=\left(1-\frac{2{\cal{M}}(R)}{R}\right)^{1/2}
\, ; \;\;\;\;\;
{\cal{M}}(0)=0
\label{eq44}
\end{equation}
is rather straightforward and, as a result, a nontrivial spherical  distribution of matter is obtained. The boundary  is usually naturally provided at the
radius
$R$ where $\rho=p=0$.
In the case when the $\rho=p=0$ point does not exist, one must integrate up to
infinity or up to a chosen cutoff radius.

It is often more convenient 
(e.g., if the equation of state is given in the form $p=p(\rho)$)
to express the field equations
in the Tolman-Oppenheimer-Volkoff (TOV) form
\begin{equation}
\frac{dp}{dr}=-(p+\rho)\frac{{\cal M}+4\pi r^3 p}{r(r-2{\cal{M}})} \, ,
\label{eq48}
\end{equation}
\begin{equation}
\frac{d{\cal{M}}}{dr}=4\pi r^2 \rho.
\label{eq49}
\end{equation}
 Here, equation (\ref{eq48}) is obtained from
 (\ref{eq40}) with the help 
of the equation of
hydrostatic equilibrium  (\ref{eq17}) which may be written as
\begin{equation}
\frac{d\ln \xi}{dr}=-\frac{1}{\rho+p}\frac{dp}{dr} \, .
\label{eq51}
\end{equation}
 The set of equations (\ref{eq48})-(\ref{eq51})  is called the
{\em TOV  equations}. 
\subsection{Newtonian Limit}
General relativity reduces to Newtonian theory in the limit of weak
gravity and low velocities. The Newtonian limit is achieved by the
approximation
$\xi=e^{\varphi}\simeq 1+\varphi$,
  ${\cal M}/r \ll 1$ , and $p\ll \rho$.
  In this limit, the two equations, (\ref{eq48}) and (\ref{eq49}),
  can be combined to give one  2nd-order differential equation  
\begin{equation}
\frac{1}{r^2}\frac{d}{dr}\frac{r^2}{\rho}\frac{dp}{dr}=
- 4\pi\rho \, .
\label{eq52}
\end{equation}
In this approximation, the relativistic energy density $\rho$ is just
 the rest mass
 density 
\begin{equation}
\rho=mn .
\label{eq53}
\end{equation}
\subsubsection{Polytropes}
If the equation of state is in the polytropic form (\ref{eq216}),
the equilibrium configurations are called {\em polytropes}.
Owing to (\ref{eq53}) we can rewrite the equation of state as
\begin{equation}
p= K \rho^{\Gamma},
\label{eq54}
\end{equation}
where 
\begin{equation}
K=\frac{\cal K}{m^{\Gamma}}\, .
\label{eq55}
\end{equation}
It is convenient to write
\begin{equation}
{\Gamma}=1+1/n,
\label{eq56}
\end{equation}
where $n$ is called the {\em polytropic index}.
Then the Newtonian equation (\ref{eq52}) can be reduced to a simple form by writing
\begin{equation}
\rho=\rho_c \theta\,; \;\;\;\;\;\; r=\alpha z ,
\label{eq57}
\end{equation}
\begin{equation}
\alpha=\left[\frac{(n+1)K \rho_c^{1/n-1}}{4\pi}\right]^{1/2} \, ,
\label{eq58}
\end{equation}
where $\rho_c$ is the central density. Then
\begin{equation}
\frac{1}{z^2}\frac{d}{dz}z^2\frac{d\theta}{dz}=-\theta^n .
\label{eq59}
\end{equation}
This is the {\em Lane-Emden equation} for the structure of a polytropic 
index $n$. This equation can be numerically
integrated starting from the center 
with the initial conditions
 \begin{equation}
\theta (0) =1; \;\;\;\; \theta'(0) =0.
\label{eq61}
\end{equation}
For $n<5$ ($\Gamma>6/5$), the solutions decrease monotonically and have a zero at a finite value $z=z_1$; $\theta(z_1)=0$. 
This point corresponds to the surface of the star, where $\rho=p=0$.
The radius is 
\begin{equation}
R=\alpha z_1=\left[\frac{(n+1)K }{4\pi}\right]^{1/2}\rho_c^{(1-n)/2n} z_1 \, ,
\label{eq62}
\end{equation}
while the mass is 
\begin{eqnarray}
M
\!&\!=\!&\!
\int4\pi r^2\rho dr 
\nonumber  \\
\!&\!=\!&\!
4\pi\alpha^3\rho_c\int_0^{z_1} z^2\theta^n dz
\nonumber  \\
\!&\!=\!&\!
-4\pi\alpha^3\rho_c\int_0^{z_1} \frac{d}{dz}z^2\frac{d\theta}{dz} dz
\nonumber  \\
\!&\!=\!&\!
4\pi\left[\frac{(n+1)K }{4\pi}\right]^{3/2}\rho_c^{(3-n)/2n} z_1^2
|\theta'_1| \, ,
\label{eq63}
\end{eqnarray}
where the absolute value $|\theta'_1|$ 
has been introduced because $\theta'_1\equiv\theta'(z_1)$ 
is negative. Eliminating $\rho_c$ between (\ref{eq62}) and (\ref{eq63})
gives the mass-radius relation for polytropes \cite{sha}:
\begin{equation}
M=4\pi R^{(3-n)/(1-n)}
\left[\frac{(n+1)K}{4\pi}\right]^{n/(n-1)}z_1^{(3-n)/(1-n)}
z_1^2|\theta'_1| \, .
\label{eq64}
\end{equation}
The solutions we are particularly interested in are
\begin{equation}
\Gamma=5/3, \;\;\;\; n=3/2,  \;\;\;\;
z_1=3.65375, \;\;\;\; 
z_1^2|\theta'_1|=2.71406\,,
\label{eq65}
\end{equation}
\begin{equation}
\Gamma=4/3, \;\;\;\; n=3,  \;\;\;\;\;\;\;
z_1=6.89685, \;\;\;\; 
z_1^2|\theta'_1|=2.01824\, ,
\label{eq66}
\end{equation}
where the values  5/3 and 4/3 of $\Gamma$ correspond to the nonrelativistic and 
ultrarelativistic regimes, respectively (see appendices 
\ref{degenerate} and \ref{polytropic}). 
\section{Compact Astrophysical Objects}
Traditionally, compact astrophysical objects represent
the final stages of  stellar evolution: 
white dwarfs (WD), neutron stars (NS), and black holes (BH).
They differ from normal stars in two basic ways.

First, since they do not burn nuclear fuel, they cannot support themselves
against gravitational collapse by generating thermal pressure.
Instead, either they are prevented from collapsing by the degeneracy pressure
(WDs and NSs) or they are completely collapsed (BHs).
With the exception of the spontaneously radiating ``mini" BHs
with masses less than $10^{12}$ kg and radii smaller than a fermi, all these
objects are essentially static over the lifetime of
the universe.

The second characteristic distinguishing compact  objects from normal stars
is their exceedingly small size. Relative to normal stars of comparable mass,
compact objects have much smaller radii and hence, much stronger surface
gravity.
\vskip .2in
\begin{tabular}{|l|cccc|} \hline
              & Mass      & Radius   & Mean Density & Surface Potential \\
 Object       &   $M$     & $R$       & g cm$^{-3}$  &   M/R     \\  \hline
 Sun          &$M_{\odot}$& $R_{\odot}$& 1           & $10^{-6}$    \\
White dwarf  &$\simlt M_{\odot}$& $\sim 10^{-2} R_{\odot}$& $\simlt 10^7$   
        & $\sim 10^{-4}$    \\  
Neutron star &$\sim 1-3 M_{\odot}$& $\sim 10^{-5} R_{\odot}$& $\simlt 10^{15}$    
        & $\sim 10^{-1}$    \\  
Black hole &Arbitrary& $ 2M$   & $\sim M/R^3$         & $\sim 1$    \\  \hline
\end{tabular}
\vskip .2in
$M_{\odot}=1.989 \times 10^{30}$ kg;
$R_{\odot}=6.9599 \times 10^5$ km

\subsection{White Dwarfs}

White dwarfs are stars that no longer burn their nuclear fuel
and their gravitational collapse is supported by the pressure of
degenerate electrons. 
Their mass and radius are about 1~$M_\odot$   and 
5000~km, respectively. 
\subsubsection{Equation of State}
We assume that the interior of a WD is
almost completely ionized plasma at a temperature
$T\ll m_e$.
Hence, the electrons are assumed to be degenerate.
The density $\rho$ of a WD is basically the density of
barionic matter (neutrons end protons).
The pressure is dominated by the pressure of electrons.
This may be seen as follows.
In the nonrelativistic regime
the pressure roughly equals the average kinetic energy
(see equation (\ref{eq314})).
Owing to the momentum conservation, the electron and proton
average momenta are equal and hence, their average kinetic energies
in the nonrelativistic regime
are inverse proportional to their masses.
Hence, the pressure of electrons is roughly by a factor
 of $m_n/m_e \simeq 2000$  larger than the pressure due to nucleons.
 
To determine the equation of state, first assume  that the electrons
are  nonrelativistic.
The  density  is related to the number density of electrons as
\begin{equation}
\rho= \eta_e m_{n}  n_e \, ,
\label{eq67}
\end{equation}
where the factor $\eta_e$,  
to a good approximation, equals the  number of nucleons 
per free electron, i.e., $\eta_e=A/Z$. For example, for completely ionized
pure $^{12}$C, $\eta_e=2$. The number of electrons is
given by (\ref{eq96}), with $m=m_e$.
Hence,
\begin{equation}
\rho=\frac{1}{3\pi^2}\eta_e m_{B} (m_e\,X)^3,
\label{eq68}
\end{equation}
where $m_e X=q_F=\sqrt{\mu^2-m_e^2}$ is the Fermi momentum
of the electrons.
We could now use the relativistic expression (\ref{eq98}) for the pressure
together with (\ref{eq68}) and numerically solve the TOV equations
to find the entire range of white-dwarf solutions. 

However, it is instructive to consider the
nonrelativistic and extreme relativistic regimes separately 
because, in these cases, the equation of state takes a polytropic form.
The pressure in the two regimes is given by (\ref{eq54}) with
$\Gamma=5/3$ for the nonrelativistic and $\Gamma =4/3$ for the extreme relativistic
regime. 
Using this and the solutions for the polytropes
(\ref{eq62})-(\ref{eq66})
we find the radius and the mass of the WD:
\begin{itemize}
\item
Low-density (nonrelativistic) regime.  
\begin{equation}
\Gamma=\frac{5}{3}; \;\;\;\;\;\;\; 
K=\frac{3^{2/3}\pi^{4/3}}{5 m_e m_n^{5/3}\eta_e^{5/3}} \, ,
\end{equation}
\begin{equation}
R= 1.122\times 10^4 \left(\frac{\rho_c}{10^6 {\rm g}\:
{\rm cm}^{-3}}\right)^{-1/6}\left(\frac{\eta_e}{2}\right)^{-5/6} {\rm km} ,
\label{eq69}
\end{equation} 
\begin{eqnarray}
M
\!&\!=\!&\!
0.4964 \left(\frac{\rho_c}{10^6 \,{\rm g}\,
{\rm cm}^{-3}}\right)^{1/2}\left(\frac{\eta_e}{2}\right)^{-5/2} M_{\odot} \, ,
\nonumber  \\
\!&\!=\!&\!
0.7011\left(\frac{R}{10^4 {\rm km}}\right)^{-3}\left(\frac{\eta_e}{2}\right)^{-5} M_{\odot}\, .
\label{eq72}
\end{eqnarray} 
\item
High-density (ultrarelativistic) regime. 
\begin{equation}
\Gamma=\frac{4}{3}; \;\;\;\;\;\;\; 
K=\frac{3^{1/3}\pi^{2/3}}{4 m_n^{4/3}\eta_e^{4/3}} \, ,
\end{equation} 
\begin{equation}
R= 3.347\times 10^4 \left(\frac{\rho_c}{10^6 \,{\rm g}\,
{\rm cm}^{-3}}\right)^{-1/3}\left(\frac{\eta_e}{2}\right)^{-2/3} {\rm km},
\label{eq74}
\end{equation}
 \begin{equation}
M= 1.457\left(\frac{\eta_e}{2}\right)^2 M_{\odot} \, .
\label{eq75}
\end{equation}
 \end{itemize}
 
 Note that $M$ is independent of $\rho_c$ in the extreme relativistic
 limit. We conclude that as $\rho_c \rightarrow \infty$, the electrons become more
 and more relativistic and the mass asymptotically approaches the value (\ref{eq75})
 as $R\rightarrow 0$.
 This mass limit is called the {\em Chandrasekhar limit},
  and represents the maximum possible mass of a WD
\cite{chand}.
  
  Of course, the limiting radius  is  not zero. The integration of the TOV equations
  using the exact degenerate Fermi gas equation of state would also
   give the limiting
  value of $R$ of the order of
   \begin{equation}
R_{\rm Ch} \sim \frac{m_{\rm Pl}}{m_n m_e}\sim 5 \times 10^3 {\rm km} .
\label{eq76}
\end{equation}

\subsection{Neutron Stars}
\label{neu}
If the mass of the collapsing star is larger than the
Chandrasekhar limit, the degeneracy pressure of the electrons can no longer 
support the gravitational attraction and the collapse does not stop. As the density
increases, the Fermi energy $E_{\rm F}=\mu$
of the electrons increases 
according to (\ref{eq68}). 

At a density of about
$2\times 10^7$ g cm$^{-3}$, 
the Fermi energy of the electrons has risen to $m_n -m_p =1.29$ MeV where
electrons can now be absorbed by protons through 
the inverse $\beta$ decay
 \begin{equation}
e^- +p \rightarrow n +\nu_e
\end{equation}
This reaction cannot come to equilibrium with the reverse reaction
\begin{equation}
n +\nu_e \rightarrow  e^- +p 
\end{equation}
because the neutrinos escape from the star and the normal $\beta$ decay
cannot occur because all electron energy levels below $E=m_n-m_p$
are occupied when
$E_{\rm F}>m_n-m_p$.
At a density in the range  $10^7\leq \rho \leq 4\times 10^{11}$ g cm$^{-3}$ 
the medium is a composition of 
separated nuclei in  equilibrium with a relativistic
electron gas.
At $ \rho \sim 4\times 10^{11}$ the ratio $n/p$ reaches a critical level.
Any further increase leads to a ``neutron drip" -- that is , a two-phase system
in which electrons, nuclei, and free neutrons coexist.
When the density exceeds  about $4 \times 10^{12}$ g cm$^{-3}$, 
more pressure is provided by neutrons than by electrons.
The neutron gas so controls the situation; one can describe  the  medium
as one vast nucleus with lower-than-normal nuclear density.
 
As the  density reaches the normal nuclear density of about
\begin{equation}
\rho_{\rm nucl}= \frac{1GeV}{(2 \,{\rm fm})^3}=
2\times 10^{14}\, {\rm g}\,{\rm cm}^{-3},
\end{equation}
there is a phase transition in which  nuclei dissolve.
The resulting fluid consists mostly of neutrons with a small ($\simlt 5\%$) fraction of 
electrons and protons. The pressure is dominated by the degenerate ($T=0, \mu_n > m_n$) gas of neutrons.
Can the degeneracy pressure of neutrons support the star against collapse?

Assuming
that the  
Newtonian calculations are still valid
at a density not very much higher than $ \rho_{\rm nucl}$,
 we can use the high-density results obtained for 
WDs but  with $m_e\rightarrow m_n$. 
The maximal mass remains the same as in (\ref{eq76}) but the 
value of the critical $R$ is of the order of
   \begin{equation}
R \sim \frac{m_{\rm Pl}}{m_n^2}\sim 2.5 {\rm km} ,
\label{eq77}
\end{equation}
which is close to the Schwarzschild radius of the sun,
so the neglect of GR effects is not justified.

\begin{figure}
\begin{center}
\includegraphics[width=.5\textwidth,trim= 0 3cm 0 4cm]{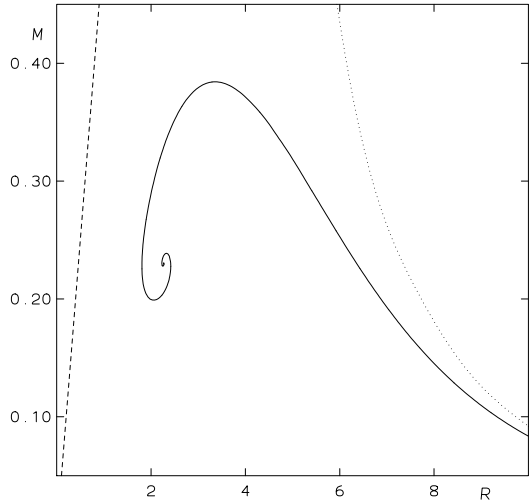}
\caption{Mass versus radius for fermion stars at
zero temperature
in the general-relativistic framework
 (solid line)
 compared with the corresponding
Newtonian approximation (dotted line).
The dashed line is the Schwarzschild BH limit $M=R/2$.}
\label{fig1}
\end{center}
\end{figure}

We have to solve the TOV equations with the equation of states of 
nuclear matter. The first numerical calculation was performed by
Oppenheimer and Volkoff \cite{opp} who used the relativistic degenerate
Fermi gas described by (\ref{eq96})-(\ref{eq98}).
In Fig.\ \ref{fig1} we plot the mass of the NS as
a function of the radius $R$. The maximum of the curve
corresponds to the
OV limit.
The limiting values are \cite{bil3}
\begin{equation}
R_{\rm OV} = 3.357 \,
\frac{m_{\rm Pl}}{m^2} \,
\left( \frac{2}{g} \right)^{1/2}
=  9.6 
\left( \frac{1 \; \mbox{GeV}}{m^{2}}
\right)^{2} \, \left( \frac{2}{g} \mbox{km}
\right)^{1/2}  ,
\label{eq900}
\end{equation}
\begin{equation}
M_{\rm OV} = 0.38426 \,
\frac{m_{\rm Pl}^3}{m^2} \,
\left( \frac{2}{g} \right)^{1/2}
=  0.7   \,
\left( \frac{1 \; \mbox{GeV}}{m^{2}}
\right)^{2} \, \left( \frac{2}{g}
\right)^{1/2}  M_{\odot},
\label{eq901}
\end{equation}
where $g$ is the fermion degeneracy factor.
This limit is reached when the central density becomes 
$\rho_c=5\times 10^{15}\, {\rm g}\,{\rm cm}^{-3}$
with $g=2$ for neutrons.
The part of the curve
left from the maximum  represents unstable configurations
that curl up around the point corresponding to the
infinite central-density limit.

However, the degenerate neutron gas equation of state is not realistic
because at such large densities the effects of nuclear forces must be included.
More realistic equations of state predict the maximum NS mass in the range
1.5 - 2.7 $M_{\odot}$. Hence, the maximal mass is rather sensitive to 
the not very well-known equation of state for nuclear matter.
Since the density inside the star varies from very large central values to
zero at the surface, the equation of state is actually rather complicated as the
star may contain different phases of nuclear matter.

The possibility to identify some compact objects as BHs relies in part
on being able to state categorically that the observed object has a mass larger than 
the maximum allowed mass of a stable NS. It turns out that it is possible
to set an upper limit to the NS mass based on  rather general
assumptions (Rhodes and Ruffini \cite{rho}):
\begin{enumerate}
\item
The TOV equation determines the equilibrium structure.
\item
The equation of state satisfies the local  stability condition
\begin{equation}
c_s^2\equiv \frac{\partial p}{\partial \rho}\geq 0,
\label{eq79}
\end{equation}
that is, the speed of sound is real. If this condition were violated,
small elements of matter would spontaneously collapse.
\item
The equation of state satisfies the causality condition
\begin{equation}
c_s^2\leq 1,
\label{eq81}
\end{equation}
that is, the speed of sound is less than the speed of light.
\item
The equation of state below some ``matching density" $\rho_0$ is known.
\end{enumerate}

Rhodes and Ruffini performed a variational calculation to determine 
which equation of state above $\rho_0$ maximizes the mass.
Then,  the numerical integration of the TOV equations for a chosen
equation of state below $\rho_0$ 
gives
\begin{equation}
M_{\rm max} \simeq 3.2 \left(\frac{\rho_0}{4.6 \times 10^{14} 
{\rm g}\,{\rm cm}^{-3}}\right)^{-1/2} M_{\odot}\, .
\label{eq82}
\end{equation}

A semianalytic treatment of Nauenberg and Chapline \cite{nau}
with similar assumptions about the equation of state gives 
\begin{equation}
M_{\rm max} \simeq 3.6  M_{\odot}\, .
\label{eq83}
\end{equation}
Abandoning the causality constraint still leads to a severe mass limit,
  assuming general relativity to be valid. 
One finds (see, e.g., Hartle and Sabbadini \cite{har})
\begin{equation}
M_{\rm max} \simeq 5.2  M_{\odot}\, .
\label{eq84}
\end{equation}
 
In their estimate, Rhodes and Ruffini took the the so-called Harrison-Wheeler
equation of state which accurately describes the nuclear matter densities
below the neutron drip.
 In fact, 
it turns out that the upper  mass limit is not very sensitive 
to the equation of state
used below $\rho_0$.
It is important to note that even if  a new 
physics exists
at subnuclear and subquark level
(preons, pre-preons etc.; see, e.g., a black-hole sceptical paper 
 \cite{han}), it is reasonable to expect that 
the equation of state will still 
satisfy the above conditions and hence, the above limits 
cannot be significantly altered.

\subsection{Black Holes}
\label{black}
What happens if the mass of the collapsing star is larger than
 the maximal allowed NS mass? 
In such a case, there is nothing to prevent the star from further collapse 
ending in a BH
\footnote{It has been recently  proposed that loop quantum gravity effects
stop the collapse to a singularity by a bounce of the infalling matter \cite{boj}}.

Astrophysical BHs  are macroscopic objects with masses ranging from
several $M_\odot$ (X-ray binaries) to $10^6-10^{9.5}M_\odot$ (in galactic
nuclei). Being so massive, these BHs are described completely by
classical general relativity.  As such, each BH is characterized by
just three numbers: mass $M$, spin parameter $a$, defined such that
the angular momentum of the BH is $J=aM$, and electric charge $Q$
 \cite{mis,sha,wei}
 (see also J.~Zanelli's lectures at
 this School).
Actually, an astrophysical BH is not likely to have any
significant electric charge because it will usually be rapidly
neutralized by surrounding plasma.  Therefore, the BH can be fully
characterized by measuring just two parameters, $M$ and $a$, of
which the latter is constrained to lie in the range from 0 (nonrotating
BH) to $M$ (maximally-rotating BH).

\subsubsection{Spherical Collapse}
\begin{figure}
\begin{center}
\input{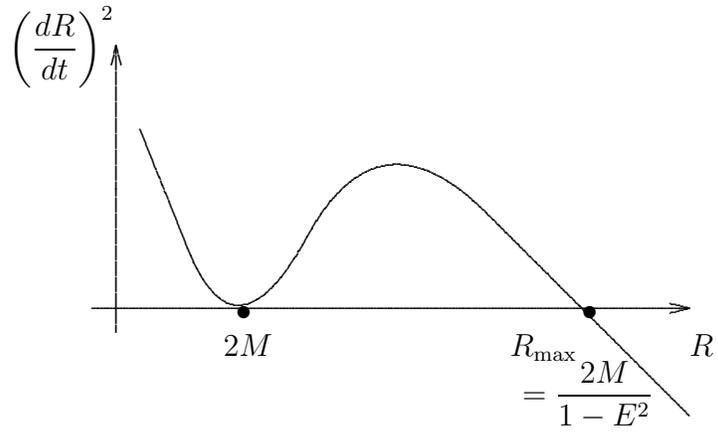}
\caption{
The surface radial velocity  squared  versus $R$
for a spherically symmetric collapse of a ball of dust
(figure from \cite{tau}). 
}
\label{dust1}
\end{center}
\end{figure}

A useful toy model that illustrates the collapse is a 
self-gravitating spherically symmetric
ball of dust (i.e., zero pressure fluid). 
Birkhoff's theorem (section \ref{bir}) implies that the metric outside the star is
Schwarzschild  (section \ref{sch}, equation (\ref{eq47})).
This is valid outside the star but also, by continuity of the metric, on the 
surface.  If $r=R(t)$ on the surface, we have 
\begin{equation}
ds^2=\left[ \left(1-\frac{2M}{R}\right)-\left(1-\frac{2M}{R}\right)^{-1}
\left(\frac{dR}{dt}\right)^2\right]dt^2-R^2(d\theta^2+\sin^2 \theta d\phi^2).  
\end{equation}
Zero pressure and spherical symmetry imply that a point on 
the surface follows a radial timelike geodesic, $d\theta=d\phi=0$ and
$ds^2=d\tau^2 > 0$, so
\begin{equation}
1=\left[\left(1-\frac{2M}{R}\right)-\left(1-\frac{2M}{R}\right)^{-1}\left(\frac{dR}{dt}\right)^2\right]
\dot{t}{\,}^2 ,
\label{eq86}
\end{equation}
where 
\begin{equation}
\dot{t}=\frac{dt}{d\tau}.
\label{eq85}
\end{equation}
Since $\xi=\partial/\partial t$ is a Killing vector,
by Proposition \ref{killing} in appendix \ref{constants}  
the energy per unit mass 
\begin{equation}
E= \xi_{\mu} \frac{dx^\mu}{d\tau} = g_{00}\dot{t}=\left(1-\frac{2M}{R}\right)\dot{t}\, . 
\label{eq85a}
\end{equation}
is a constant of motion.
Note that $E>0$ for timelike and null geodesics as long
as $\xi^\mu$ is timelike, i.e., as long as
$R>2M$. The quantity 
$E$ is constant along the geodesics and $E < 1$ for 
gravitationally bound particles.  Using  (\ref{eq85a}) in 
(\ref{eq86}) gives
\begin{equation}
\left(\frac{dR}{dt}\right)^2=\frac{1}{E^2}\left(1-\frac{2M}{R}\right)^2\left(\frac{2M}{R}-1+
E^2\right).
\label{eq87}
\end{equation}
We plot this function in Fig.\ \ref{dust1}.
The surface radial velocity  $dR/dt$ as a function of $R$ 
has a zero at $R=R_{\rm max}$ and a minimum at $R=2M$. 
We consider the collapse to begin at $R=R_{\rm max}$ with 
zero velocity. The radius $R$ then decreases and approaches $R=2M$
asymptotically as $t\rightarrow \infty$. 
This may be seen by integrating
\begin{equation}
t=E\int_{R_{\rm max}}^{R_{\rm min}} dR (1-2M/R)^{-1}
(2M/R-1+E^2)^{-1/2},
\label{eq88}
\end{equation}
where $R_{\rm min}>2M$. 
Obviously $t\rightarrow \infty$ as $R_{\rm min}\rightarrow 2M$.
 So an observer ``sees'' the star contract at most
to $R=2M$ but no further. 

However, from the point of view of an observer on the surface of the star, the 
relevant time variable is the proper time along a radial
geodesic, so we use
\begin{equation}
\frac{d}{dt}=\frac{1}{\dot{t}}\frac{d}{d\tau}=
\frac{1}{E}\left(1-\frac{2M}{R}\right)\frac{d}{d\tau}
\end{equation}
to rewrite (\ref{eq87}) as
\begin{equation}
\left(\frac{dR}{d\tau}\right)^2=\left(\frac{2M}{R}-1+E^2\right)=
(1-E^2)\left(\frac{R_{\rm max}}{R}-1\right).
\end{equation}
\begin{figure}
\begin{center}
\input{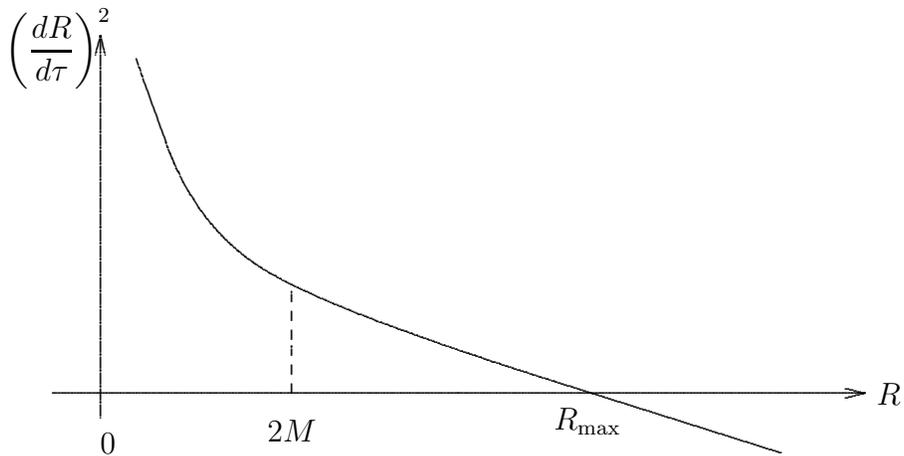}
\caption{
The surface radial velocity  as seen by the observer 
on the surface
(figure from \cite{tau}). 
}
\label{dust2}
\end{center}
\end{figure}
The  star surface falls from $R=R_{\rm max}$ through $R=2M$ in 
{\bf finite proper time}.  In fact, it reaches the $R=0$ singularity 
in the proper time 
\begin{equation}
\tau=\frac{\pi M}{(1-E)^{3/2}} \, .
\end{equation}
So, for an observer following radial geodesics, 
the star collapses to a singularity in a finite proper time of the order of 
$10^{-5}$ s ($ \sim R_{\rm Sch}/c= 3{\rm km}/300\, 000\: {\rm km} \: {\rm s}^{-1} $).
 
Nothing special happens at $R=2M$, which suggests that we investigate the 
spacetime near $R=2M$ in coordinates adapted to infalling observers.
\subsubsection{Eddington-Finkelstein Coordinates}
A singularity of the metric is a point at which the determinant of either 
the metric tensor or
of its inverse vanishes, or at which some elements of the metric 
or its inverse diverge.
However, a singularity of the metric may be simply due
to a failure of the coordinate system. A simple two-dimensional example is the
origin in plane polar coordinates.  Such
singularities, usually referred to as 
{\em coordinate singularities}, are removable by a coordinate transformation.

If no coordinate system exists for which the
singularity is removable, then it is irremovable, i.e., a genuine singularity of
 spacetime. Any singularity for which some scalar constructed from the
curvature tensor blows up as it is approached, is irremovable. Such singularities
are called {\em curvature singularities} or
{\em true singularities}. The singularity at $r=0$ in the
Schwarzschild metric is an example. 

We now show that the apparent singularity of the
 Schwarzschild metric at the Schwarz\-schild radius is removable. 
On radial null geodesics in Schwarzschild spacetime
\begin{equation}
dt^2=\frac{1}{\left(1-2M/r\right)^2}dr^2\equiv\left(dr^*\right)^2 ,
\end{equation}
where
\begin{equation}
r^*=r+2M\ln\left| \frac{r-2M}{2M}\right| 
\end{equation}
is the {\em Regge-Wheeler} or {\em tortoise radial coordinate}.  
As $r$ ranges from $2M$ 
to $\infty$, $r^*$ ranges from $-\infty$ to $\infty$.  Thus 
\begin{equation}
d(t\pm r^*)=0 
\end{equation}
on radial null geodesics.
Now, define a new time coordinate (ingoing radial null coordinate) $v$ by 
\begin{equation}
v=t+r^*,\;\;\;\;\;  -\infty<v<\infty
\end{equation}
and rewrite the Schwarzschild metric in {\em ingoing Eddington-Finkelstein 
coordinates}
\cite{edd}
$v$, $r$, $\theta$, and $\phi$:
\begin{eqnarray}
ds^2 
\; & \;= \; & \;
\left(1-\frac{2M}{r}\right)\left(dt^2-d{r^*}^2\right)-r^2d\Omega^2 
 \nonumber \\
 \; & \;= \; & \;
 \left(1-\frac{2M}{r}\right)dv^2-2dr\,dv-r^2d\Omega^2 .
\end{eqnarray}
This metric is  initially defined for $r>2M$ because the relation 
$v=t+r^*(r)$ between $v$ and $r$ is defined only for $r>2M$, but it can now be
analytically continued to all $r>0$.  Because of the $dr\,dv$ cross-term,
the metric in EF coordinates is  nonsingular at $r=2M$, so the
singularity in Schwarzschild coordinates is really a coordinate singularity. 
There is nothing at $r=2M$ to prevent the star from collapsing through $r=2M$. 
 
However,
  no future-directed 
timelike or null worldline can reach $r>2M$ from $r\le 2M$.
This may be seen as follows.
 When $r\le 2M$, we have
\begin{equation}
2dr\,dv  = -ds^2 - \left(\frac{2M}{r}-1\right)dv^2- r^2d\Omega^2 
   \le 0 \;\;\; \mbox{when}\; ds^2\ge 0 \, .
\end{equation}
Hence, for all timelike or 
null worldlines, $dr\,dv\le 0$. The null coordinate $dv>0$ for future-directed 
worldlines, so $dr\le 0$ with equality when $r=2M$, $d\Omega=0$ (i.e., ingoing
radial null geodesics at $r=2M$). 
 
Thus, no signal from the star's surface can escape to infinity once the surface has 
passed through $r=2M$.  The star has collapsed to a {\em black hole}.  For
the external observer, the surface never actually reaches $r=2M$, but as $r\to
2M$, the redshift of light leaving the surface increases exponentially 
fast \cite{wei} and the star effectively disappears from view within a time 
$\sim M/m_{\rm Pl}^2 \simeq 5 \times 10^{-6}$ s . 
The late time appearance is dominated by photons escaping from the unstable
photon orbit at $r=3M$.

From the point of view of an outside observer, a star collapsing 
to a black hole never appears to collapse, but rather freezes at the horizon. 
How then can it be said that the star collapses to a singularity, 
if it never appears to collapse even till the end of the Universe?

The star does in fact collapse inside the horizon, even though 
an outside observer sees the star freezing at the horizon. The freezing can be
regarded as a light travel time effect. Space can be regarded as falling 
into the black hole, reaching the speed of light at the horizon, 
and exceeding the speed of light inside the horizon
\cite{river}. 
The photons that are exactly at the horizon and are pointed 
radially upwards stay there for ever, 
their outward motion through space at the speed of light being 
canceled by the inward flow of space at the speed of light. 
It follows that it takes an infinite time for light to travel
 from the horizon to the external observer. The star does actually collapse:
  it just takes an infinite time for the information that it has
   collapsed to get to the outside world!

\section{Rotating Black Holes}
 The spacetime around a rotating black hole is described by the Kerr
metric expressed in Boyer-Lindquist coordinates  \cite{mis,sha,wal} 
 \begin{eqnarray}
 ds^2 
 & = & 
 \left(1- \frac{2Mr}{\Sigma}\right)dt^2 
 + \frac{4aMr \sin^2\theta }{\Sigma}dt\,d\phi
\nonumber \\
 & & 
 -\left( \frac{ \left(r^2+a^2\right)^2-\Delta a^2\sin^2\theta}{\Sigma}
\right)\sin^2\theta d\phi^2 
-\frac{\Sigma}{\Delta}dr^2-\Sigma d\theta^2,
\label{eq902} 
\end{eqnarray}  
with
\begin{equation}
\Delta = r^2-2Mr+a^2 , \;\;\;\;\;\;\nonumber \\
\Sigma  =  r^2+a^2\cos^2\theta \, .  
\label{eq903} 
\end{equation}  
The parameters  $a$ and $M$ are related to the total angular
momentum $J=aM$.
The horizon occurs at those points where $\Delta=0$, i.e., at
the roots of the quadratic equation $\Delta=0$
\begin{equation}
 r_{\pm}=M\pm \sqrt{M^2-a^2}\, .
\label{eq914} 
\end{equation} 
Note that $|a|$  must be less than $M$ for a black hole
to exist. If $a$ exceeded $M$, one would have a gravitational field
with a ``naked" singularity, i.e., one not within an event horizon.
A naked singularity is forbidden by the so called 
{\em cosmic censor conjecture} (for details, see Wald \cite{wal} and references
therein).
A black hole with $|a|=M$ is called a maximally rotating black hole.
\vskip \baselineskip
\noindent
{\bf Remarks}
\begin{itemize}
\item When $a=0$, the Kerr solution reduces to the Schwarzschild solution.
\item Taking $\phi \to -\phi$ effectively changes the sign of $a$, so we 
may choose $a \ge 0$ without loss of generality.
\end{itemize} 
\subsection{Geodesic Motion}
A straightforward approach to finding all geodesics is to
integrate  the geodesic equation (\ref{eq204}) directly.
However, it is often more economic to simplify the problem
by symmetry considerations.
The  Kerr spacetime is stationary and axially symmetric so
there exist two Killing vectors
$\xi$ and $\psi$ which by Proposition \ref{killing}
(appendix \ref{constants}) yield
a conserved energy $E$ and an angular momentum $L$ (per unit mass)
along geodesics
\begin{equation}
 E= u^\mu\xi_\mu= \left(1-\frac{2Mr}{\Sigma}\right) \dot{t}
 +\frac{2Mar \sin^2\theta}{\Sigma}\dot{\phi} \, ,
\label{eq905} 
\end{equation} 
\begin{equation}
 L= -u^\mu\psi_\mu= 
- \frac{2Mar \sin^2\theta}{\Sigma}\dot{t}
+\left( \frac{ \left(r^2+a^2\right)^2-\Delta a^2\sin^2\theta}{\Sigma}
\right)\sin^2\theta \dot{\phi} \, ,
\label{eq906} 
\end{equation} 
where $\dot{x}^\mu=u^\mu=dx^\mu/d\tau$. In addition we have 
\begin{equation}
g_{\mu\nu} \dot{x}^\mu\dot{x}^\nu =\kappa ,
\label{eq907} 
\end{equation} 
where $\kappa =$ 1, 0, and $-1$ for timelike, null, and spacelike geodesics,
respectively.
One may use equations (\ref{eq905}) and (\ref{eq906}) to solve for
$ \dot{t}$ and $\dot{\phi}$ in terms of $E$ and $L$, and 
substituting the results into (\ref{eq907}) one obtains a
differential equation for $\dot{r}$. In the case of 
equatorial geodesics, $\theta=\pi/2$, one finds
\begin{equation}
\frac{1}{2}\dot{r}+ V(r)=0 ,
\label{eq908} 
\end{equation} 
where
\begin{equation}
V=-\kappa\frac{M}{r}+\frac{L^2}{2r^2}+\frac{1}{2}
(\kappa-E^2)\left(1+\frac{a^2}{r^2}\right)
-\frac{M}{r^3}(L-aE)^2 .
\label{eq909} 
\end{equation} 
Thus, the problem of obtaining the  geodesics in the
equatorial plane reduces to solving a problem of ordinary,
nonrelativistic, one-dimensional motion in an effective potential.
The calculations are relatively simple for circular orbits.
Circular orbits occur where $\dot{r}=0$, which requires
\begin{equation}
V=0, \;\;\;\;\;\;\; \frac{\partial V}{\partial r}=0 .
\label{eq910} 
\end{equation} 
For $\kappa>0$, equations (\ref{eq910}) have  solutions
$\tilde{E}(r)$ and $\tilde{L}(r)$
for all  $r>r_{\rm ph}$ (for details, see \cite{sha,bar}),
where $r>r_{\rm ph}$ is the radius of the photon circular orbit.

\subsubsection{Photon Circular Orbit}
Photon circular orbits are possible only for 
particular radii $r_{\rm ph}$.
The requirements (\ref{eq910}) 
for null geodesics ($\kappa=0$ in (\ref{eq909}))
yield a cubic equation in $\sqrt{r}$
 \begin{equation}
r^2-2M \pm 2a \sqrt{Mr} =0
\label{eq912} 
\end{equation}
with the solution \cite{bar} 
\begin{equation}
r_{\rm ph}=2M\left\{1+ \cos \left[\frac{2}{3} \cos^{-1}(\mp a/M)\right]\right\} .
\label{eq913} 
\end{equation} 
For $a=0$, $r_{\rm ph}=3M$, while for $a=M$, $r_{\rm ph}=M$ (corotating) 
or $r_{\rm ph}=4M$ (counterrotating orbit).
For $\kappa=0$, equations (\ref{eq910}) have a solution
 \begin{equation}
L=\pm E \sqrt{3 r_{\rm ph}^2 + a^2}\, .
\label{eq911} 
\end{equation} 
The photon orbit is the innermost boundary of the circular orbits for particles,
i.e., for timelike geodesics ($\kappa>0$).

\subsubsection{Angular Velocity}
\label{angular}
The angular velocity of an orbiting particle is defined as
\begin{equation}
\Omega=\frac{u^\phi}{u^0}=\frac{\dot{\phi}}{\dot{t}} \, .
\label{eq930} 
\end{equation} 
Using the parameterization (\ref{eq031}) for the velocity components, 
it may be easily shown (exercise) that
 $\Omega$ 
can be  expressed as
\begin{equation}
\Omega= -\frac{g_{0\phi}+\lambda g_{00}}{g_{\phi\phi}+\lambda g_{0\phi}} \, ,
\end{equation}
where
\begin{equation}
\lambda = -\frac{L}{E}=-\frac{u_{\phi}}{u_0}\, .
\end{equation}
For equatorial circular orbits one finds \cite{bar}
\begin{equation}
\Omega=\pm\frac{M^{1/2}}{r^{3/2}\pm aM^{1/2}}.
\label{eq931} 
\end{equation} 
This is the general-relativistic form of Kepler's third law
for equatorial circular orbits. In this case, the quantity 
$\Omega$ is called the {\em Keplerian frequency}.

\subsubsection{Innermost Stable Circular Orbit}
\label{innermost}
It is obvious that not all circular orbits will be stable because,
in addition to 
(\ref{eq910}), stability requires
\begin{equation}
\frac{\partial^2 V}{\partial r^2}\geq 0.
\label{eq915} 
\end{equation}
 From (\ref{eq909}) we obtain 
\begin{equation}
1-E^2 \geq \frac{2M}{3r}\, .
\label{eq916} 
\end{equation}
Substituting the solution $\tilde{E}(r)$ for timelike geodesics,
we obtain a quartic equation in $\sqrt{r}$ for the limiting case of equality.
The solution $r_{\rm is}$ is the radius of the {\em innermost stable 
circular orbit} (ISCO) also referred to 
as the {\em marginally stable orbit} \cite{bar}
\begin{equation}
r_{\rm is}=
M \left[3+Z_2\mp (3-Z_1)^{1/2}(3+Z_1+2Z_2)^{1/2}\right],
\label{eq917} 
\end{equation}
with
 \begin{eqnarray}
Z_1
\!&\!=\!&
1+\left(1-\frac{a^2}{M^2}\right)^{1/3}
\left[\left(1+\frac{a^2}{M^2}\right)^{1/3}+
\left(1-\frac{a^2}{M^2}\right)^{1/3}
 \right],
\nonumber \\
Z_2
\!&\!=\!&
\left(3\frac{a^2}{M^2}+Z_1^2 \right)^{1/2}.
\end{eqnarray}
For $a=0$, $r_{\rm is}=6M$; for $a=M$, $r_{\rm is}=M$ (corotating)
$r_{\rm is}=9M$ (counterrotating).
Obviously, compared with the photon circular orbit, the 
innermost stable circular orbit satisfies $r_{\rm is}\geq r_{\rm ph}$.

A quantity of great interest for the potential efficiency of a BH accretion disk
as an energy source is the binding energy of the ISCO.
Defining the efficiency $\eta$ as the maximum binding energy per unit rest mass, from 
 (\ref{eq916})  (with =)
one finds 
 \begin{equation}
\eta\equiv 1- \tilde{E}_{\rm is}=1-
\left(1-\frac{2M}{3r_{\rm is}}\right)^{1/2}.
\label{eq918} 
\end{equation}
Plugging in the solution (\ref{eq917}), one finds that
the efficiency $\eta$ increases from
$1-\sqrt{8/9}$ (a=0) to $1-\sqrt{1/3}$ (a=M) for corotating orbits, while
it decreases from $1-\sqrt{8/9}$ (a=0) to $1-\sqrt{25/27}$ (a=M)
for counterrotating orbits. The maximum binding energy for a maximally
rotating BH is 42.3\% of the rest-mass! This is the amount of energy that
is released by matter spiraling in toward the BH through a succession
of almost circular equatorial orbits. A negligible amount of energy
is released during the final plunge from $r_{\rm is}$ into the BH.

\subsection{Ergosphere} 
\label{ergo}
A curious  property of rotating BHs is that
there exist  particle trajectories (i.e., timelike geodesics)
with negative energies. 
The energy defined by (\ref{eq905}) can  be negative only if 
the time translation Killing vector $\xi$ is spacelike.  
The  vector $\xi$ is normally timelike at $\infty$ and
in Schwarzschild spacetime it is  timelike everywhere outside the
horizon.
However, in Kerr spacetime it need not be timelike everywhere 
outside the horizon because  
\begin{equation}
\xi^\mu\xi_\mu=g_{00}=
\left( 1-\frac{ 2Mr }{r^2+a^2\cos^2\theta }\right),
\end{equation}
so $\xi$ is timelike only if
\begin{equation}
r^2+a^2\cos^2\theta -2Mr > 0 .
\end{equation}
This implies 
\begin{eqnarray}
r
\! & \! < \! & 
  M-\sqrt{M^2-a^2\cos^2\theta} ,   
  \nonumber \\
  r
\! & \! > \! &   
  M+\sqrt{M^2-a^2\cos^2\theta} .
\end{eqnarray}
The inner boundary of this region is not physically relevant 
as it is beyond the  horizon.
The outer  boundary , i.e.,the hypersurface 
\begin{equation}
r=M+\sqrt{M^2-a^2\cos^2\theta}
\end{equation}
is called the {\em ergosphere} or 
the {\em stationary limit surface}. 
The ergosphere intersects the event horizon at 
$\theta=0,\pi$, but it lies {\bf outside} the horizon for other values of
$\theta$.  Thus, $\xi$ can become spacelike in a region outside the event
horizon.  This region is called the {\em ergoregion}. 
Hence, a particle trajectory inside the ergosphere may have negative energy!

\subsubsection{Penrose Process}
\label{penrose}

In 1969 Penrose exploited this property of Kerr BHs to design a mechanism
for the extraction of energy from a BH
\cite{pen}.
The mechanism proposed by Penrose can be understood as follows. 
Suppose that a particle approaches a Kerr BH along a geodesic.  If 
$q^\mu$ is its 4-momentum, we can identify the constant of  motion
\begin{equation}
E=q^{\mu}\xi_\mu
\end{equation}
as its energy (see section \ref{constants}). 
The trajectory is chosen so that it penetrates the ergosphere. Now suppose that the particle 
decays into two others, one of which falls into the hole
with energy $E_{\rm in}$,  while the other escapes
to $\infty$ with energy $E_{\rm out}$.
By conservation of energy
\begin{equation}
E_{\rm out}=E-E_{\rm in} \, .
\end{equation}
Normally, $E_{\rm in}>0$, so $E_{\rm out} < E$, 
but $E_{\rm in}=q_{\rm in}^{\mu}\xi_\mu$
is  not necessarily positive in the ergoregion because $\xi$ may 
be spacelike there.  Thus, if the decay takes place in the ergoregion, we may
have $E_{\rm out}>E$, so {\bf energy has been extracted from the black hole}.

The energy extraction by the Penrose process is limited by 
the area theorem of BH mechanics, which states that the surface area of the
BH horizon never decreases \cite{haw}. 
The area of the horizon is
\begin{equation}
A= \int \sqrt{ \det g}\: d\theta d\phi=
8\pi M \left(M+\sqrt{M^2-a^2} \right),
\label{eq919}
\end{equation}
where $\det g=g_{\theta\theta}g_{\phi\phi}$ is the 
determinant of the metric on the horizon surface. 
The surface metric  is obtained from (\ref{eq902})
by setting
$dt=dr=0$, $r=r_{+}=M+\sqrt{M^2-a^2}$.
The maximum energy extracted by the Penrose process is obtained if
the BH spin reduces to zero, i.e., if the BH becomes Schwarzschild.
A Schwarzschild BH with the same area will have a mass
 $M_{\rm irr}$ (irreducible mass) which satisfies
\begin{equation}
16\pi M_{\rm irr}^2=8\pi 
M \left(M+\sqrt{M^2-a^2} \right).
\label{eq920}
\end{equation}
This gives $M_{\rm irr}^2=M/2$ for a maximally rotating BH ($a=M$). 
Hence, the maximally extracted energy from a BH of mass $M$ is
$M-M_{\rm irr}= M(1-1/\sqrt{2})$. This represents $\approx 29$\%
of the mass energy of the BH. 

The Penrose process for extracting energy from a rotating BH
requires particular conditions \cite{bar} which are very difficult to
realize in nature. A more promising mechanism for extracting energy 
from a rotating BH is via magnetic fields   
(Znajek and Blandford \cite{zna,bla},) because magnetic fields are capable of connecting
regions very close to the BH to regions farther out.
Recent general-relativistic magneto-hydrodynamics (MHD)
simulations  of a magnetized plasma in the ergosphere of 
a rotating BH (Koide et al. \cite{koi1,nis,koi2}, Semenov et al. \cite{sem},
de Villiers et al. \cite{dev}, McKinney and Gammie \cite{mck})
show that, near the equatorial plane, the field lines are azimuthally twisted
and the twist then propagates outward and transports the energy along the rotation 
axis. The process is dubbed the {\em MHD Penrose process} because of a close
analogy between this mechanism \cite{koi2} and Penrose's original idea \cite{pen}.

\section{Stellar Mass BHs versus NSs}

Astrophysical observation of NSs and BHs is difficult because they are
dark and very compact.
Besides, a BH the mass of which is of the order of a few  solar masses is not 
easily distingushed from a NS because of their similar properties.
So far, NSs and stellar mass BH candidates have been identified only as the so-called
{\em X-ray binaries}. 

\subsection{X-ray Binaries}
\label{xray}
An  X-ray binary is  an  X-ray source
with an optical companion, usually a normal star.
Most of the Galactic X-ray sources are probably compact objects
accreting gas from the companion star.

This interpretation of the observational data follows from
these facts \cite{sha}:
\begin{enumerate}
\item
The variability of  X-ray emission on short timescales 
implies a small emitting region.
\item
Many of the sources are positively confirmed to be in binary systems, with
optical primaries orbiting optically invisible secondaries.
\item
 Mass  accretion onto a compact object, 
 especially a neutron star or a black hole, 
is an extremely efficient means of converting released gravitational 
potential energy into X-ray radiation.
\end{enumerate}

In general, the list of possible Galactic X-ray source candidates
includes all three kinds of compact objects:
WDs, NSs, and BHs. But in special cases, the specific nature of the 
compact object can be identified.

\subsubsection{Binary X-ray Pulsars} 
Binary X-ray sources displaying {\em periodic} 
variations are called  {\em binary X-ray pulsars}.
The pulse  periods are observed in the range $0.5 \:{\rm s}\simlt P\simlt 1000$~s.
Those with short periods  of about 1 s  are
normally identified with rotating NS. 

The standard model explains X-ray emission as due to the conversion
of the kinetic energy of the accreting matter (coming from the intense
stellar wind of the companion optical star) into radiation, because of the
interactions with the strong 
magnetic field\footnote{Obtained from conservation of the
magnetic flux during the process of collapse from a ``normal'' star
($B\sim 10^{-3}$-$10^{-2}$~T, $R\sim 10^6$~km) to a neutron star ($R\sim 10$~km)
} 
of the neutron star, of the
order of $10^7$--$10^9$~T.
The magnetic field of the compact object drives the accreted
matter onto the magnetic polar caps, and if the magnetic field axis is not
aligned with the spin axis, then the compact object acts as a
``lighthouse'', giving rise to pulsed emission when the beam (or the
beams, according to the geometry) crosses our line of sight. 

The reasons why WD and BHs cannot be pulsars are:
\begin{itemize}
\item
Rotating WD are excluded because they are too large and their density
too low to explain such short periods.
\item
Rotating BH are excluded because they are axially symmetric and have no structure
on which to attach a periodic emitter. 
Any mechanism depending on accretion
 would not be periodic to the observed precision.
 However, accretion disks round black holes could produce
 the so-called {\em quasi periodic oscillations} (see section \ref{spin})

\end{itemize}
\subsubsection{Mass Measurements}
\label{mass}
The first important criteria that may be used to distinguish a BH from a NS
are their masses. 
The fact that there exists a
 maximum mass of
a compact relativistic star such as a NS of the order $\sim 3M_\odot$
(see  section.  \ref{neu}) allows the following 
simple criteria for identifying BH candidates
\cite{nar}:

\noindent
{\em If  a compact  astrophysical object has the mass larger
 than about $3M_\odot$, then the
object is very likely a black hole.}

The most reliable means of determining astronomical masses are via
Kepler's Third Law. 
Consider two spherical masses $M_1$ and $M_2$ in a circular orbit about
their center of mass (CM). The separation of the two masses is $a$ and
their distances from the CM are $a_1$ and $a_2$.
Clearly, $a=a_1+a_2$ and $M_1 a_1=M_2 a_2$ by the definition of the CM.
Any spectra emitted from, e.g.,  $M_2$ will be Doppler shifted, depending
on the orbital velocity projection $v_2$ of $M_2$ along the
line of sight:
 \begin{equation}
 v_2=\frac{2\pi}{P_{\rm orb}} a_2 \sin i \, ,
\label{eq89}
\end{equation} 
where $P_{\rm orb}$ is the orbital period and $i$ the inclination of the
orbital plane to the line of sight.
Thus, if the spectrum of $M_1$ shows periodic variations, 
then $P_{\rm orb}$ and $v_2$
can be measured and hence one gets $a_2 \sin i$.
Alternatively, for X-ray pulses one can measure periodic
variations in the time of arrival of pulses.
The amplitude of these variations is simply the light travel time across
the projected orbit -- that is, $ a_2 \sin i/c$. 

Now,  Kepler's Third Law states
 \begin{equation}
 \frac{G(M_1+M_2)}{a^3}=\left( \frac{2\pi}{P_{\rm orb}}\right)^2.
\label{eq90}
\end{equation} 
Note that this is valid also for  elliptical orbits in which case
 $a$ is the semimajor axis of the ellipse.
Using this and 
\begin{equation}
 a=\frac{M_1+M_2}{M_1}a_2 \, ,
\label{eq91}
\end{equation} 
we obtain
\begin{equation}
 f(M_1)
 \equiv \frac{(M_1 \sin i)^3}{(M_1+M_2)^2}= 
  \frac{P_{\rm orb}v_2^3}{2\pi G} .
\label{eq92}
\end{equation} 
The quantity $f$ is called the ``mass function" and depends only on the
observable quantities $P_{\rm orb}$ and $v_2$ (or $ a_2 \sin i$).

For several X-ray binaries, it has been possible to measure 
the mass functions $f_O=f(M_O)$ and $f_X=f(M_X)$ for  both the optical companion
and the X-ray source, respectively. 
The ratio $f_O/f_X$ gives th mass ratio
\begin{equation}
 q\equiv\frac{M_O}{M_X}=\left(\frac{f_O}{f_X}\right)^{1/3}
 \label{eq93}
\end{equation} 
and then  from (\ref{eq92}) we can write
\begin{equation}
 M_X=f_X\frac{(1+q)^2}{\sin^3i}\, .
 \label{eq94}
\end{equation} 
 A unique value for $M_X$ still depends on knowing $\sin i$.
 In practice,  the X-ray eclipse duration and/or 
 variation in the optical light curve 
 are used to set geometrical constraints 
 on $\sin i$.
 
 In this way
 a complete description of the binary system has been
 obtained for six eclipsing X-ray pulsars
 with optical companions \cite{sha}.
 Their masses range 
  from 1 to 2.3 $M_\odot$, with error bars such that
 all the data fit in the range 1.2 - 1.6 $M_\odot$, which is
 expected on the basis of current theoretical scenarios for 
 NS formation.
  One of these X-ray pulsars is the famous
 Hercules X-1 neutron star discovered in 1972 \cite{tan}
 in the data of the first astronomy satellite, {\em Uhuru}
 launched by NASA off the coast of Kenya.\footnote{
 ``Uhuru" means ``freedom" in Swahili; the launch occurred on the
 anniversary of Kenya's independence.}

 \subsection{Black-Hole Binaries}
 Black hole X-ray binaries, or short,
 {\em BH binaries} are the binary X-ray sources with 
 observed masses larger than 3 $M_\odot$ with 
  nonperiodic time variability.
 As of today there are a total of 20 confirmed BH binaries
 (Remillard and McClintock \cite{rem})
 Their large masses makes them
  strong candidates for BHs.
 However,
 the mass estimates are  reliable only for those for which
  the inclination angle 
 $i$ is well known, which is not always the case.
  Fortunately, 
 according to (\ref{eq94}), 
 $M > f (M)$. 
 The mass function $f(M)$, which depends
only on the two accurately measured quantities $v_2$ and $P_{\rm
orb}$, is a strict lower bound on $M$.  Most of the 20 X-ray 
binaries have
$f(M)$ itself larger than or of the order $3M_\odot$.  Therefore, these
systems are excellent BH candidates, regardless of uncertainties in
their inclinations and companion star masses.
 \subsubsection{Cygnus X-1: A Black-Hole Candidate}
Cygnus X-1 is a typical BH binary discovered in 1972 
(Webster and Murdin \cite{web},  Bolton \cite{bol}) as a first 
stellar-mass BH candidate. 
The X-ray source (X) of Cyg X-1 is variable on all timescales varying 
from ms to months and years.
Recent observations show that Cyg X-1 periodically cycles through two accretion or spectral
states: {\em hard} and {\em low} X-ray states \cite{fen}.
The most dramatic variability is the 1-ms bursts, which set a maximum size
for  X of the order $R\simlt 300$~km and establish the object
to be highly compact.

The optical companion star  (O) in the Cyg X-1 system is a typical
 supergiant star with a well-known spectrum and
 the mass of at least $ M_O\sim 8.5 M_\odot$ \cite{sha}.
From the measurements  of the orbital elements, $P_{\rm orb}= 5.6$~days 
and $a_O \sin i=(5.82\pm 0.08) \times 10^6$~km, one obtains the mass function
of X, $f_X=(0.252\pm 0.010) M_\odot$.
Setting $\sin i= 1$ in (\ref{eq94}), one obtains a minimum value for the mass 
\begin{equation}
 M_X\simgt 3.3 M_\odot\, .
 \label{eq100}
\end{equation} 

It is possible to set a convincing  lower limit 
assuming {\em nothing} about the mass of O 
and
using only the absence
of a prominent X-ray eclipse and the estimate of the radius of O 
\cite{pac}.
 The absence of eclipse implies
 \begin{equation}
 \cos i \geq \frac{R}{a}\, ,
 \label{eq104}
\end{equation} 
where $R$ is the radius of O and $a=a_X+a_O$ is the separation 
of the two objects. Using (\ref{eq91}), we can write this as
 \begin{equation}
 \cos i \geq \frac{R}{a_O \sin i}
 \frac{M_X\sin i}{M_X+ M_O}\, .
 \label{eq105}
\end{equation} 
Hence,
\begin{equation}
 M_X \sin i\cos^2 i \geq \frac{f_X R^2}{(a_O \sin i)^2}\, .
  \label{eq106}
\end{equation} 
The function $\sin x \cos x^2$ has  a maximum value of $2/(3 \sqrt{3})$,
 and so
\begin{equation}
 M_X  \geq \frac{3 \sqrt{3}f_X R^2}{2(a_O \sin i)^2}\, .
  \label{eq107}
\end{equation} 
Estimating
 $R$  from luminosity and the effective temperature,
 one finds \cite{sha}
 \begin{equation}
 R = 6.62\times 10^6 \left(\frac{d}{1\:{\rm kpc}} \right) \: {\rm km}.
 \label{eq108}
\end{equation}  
From (\ref{eq108}) with known $f_X$ and $a_O \sin i$ one finds
\begin{equation}
 M_X  \geq 3.4 \left(\frac{d}{2\:{\rm kpc}} \right)^2 M_\odot\, .
 \label{eq109}
\end{equation} 
The distance $d$ to Cyg  X-1 is determined using two methods:
a) from the assumed luminosity compared with the apparent luminosity of
O and b) from the absorption vs. distance curve
calibrated from a large sample of stars in the same direction.
The estimated distance for O is $d\simeq 2.5$~kpc with an absolute minimum 
 of 2~kpc in order to produce the observed absorption.

We  can summarize the situation as follows: the lower limit 
(\ref{eq109}) of $3.4M_\odot$ is 
very solid. Adopting the more reasonable value of $d\simeq 2.5$~kpc 
increases this to
5.3~$M_\odot$. Various other less rigorous but more realistic arguments,
as well as more recent measurements of $R$, $d$, and $f_X$  give even 
larger values. The currently accepted mass range of Cyg X-1 is
6.8-13.3~$M_\odot$ \cite{rem}.

\subsection{Spin Estimates}
\label{spin}
Measuring the BH spin amounts to measuring the Kerr-spacetime parameter $a$.
In contrast to
BH mass estimates where  Newtonian gravity applies,
the spin of a BH or of any other rotating astrophysical object does not have
any Newtonian effect on the surrounding objects.
Only for relativistic orbits does spin have measurable effects.
Therefore, to measure $a$, we need test particles orbiting very close
to the innermost stable circular orbit (ISCO) 
(see section \ref{innermost}).
Such test particles are provided by the accretion disk.

The gas in an accretion disk starts from large radii and spirals in
through a sequence of nearly circular orbits as it viscously loses
angular momentum.
The main source of instability and loss of angular momentum
are  MHD effects, e.g.,
magnetorotational instability \cite{bal1} and
magnetoviscous instability \cite{bal2}.
  When the gas reaches the ISCO, no more stable
circular orbits are available, so the gas accelerates radially and
free-falls into the BH.  Thus, the ISCO serves effectively as the
inner edge\footnote{The inner edge may not be
  very pronounced; compare, e.g., the truncated disk model
  \cite{mul}}
  of the accretion disk.  A variety of observational methods
have been proposed for estimating the radius $r_{\rm in}$ 
(spectral fitting method; relativistic iron line method),
the Keplerian frequency $\Omega$ (quasiperiodic oscillations method),
or the binding energy $\eta$ at $r_{\rm is}$.
Next, we briefly describe some of these  methods.
For more details and for  references, see  \cite{nar}.
\vskip \baselineskip
\noindent
{\bf Spectral Fitting}
\vskip \baselineskip
When a BH has a large mass accretion rate $\dot M$, corresponding to
an accretion luminosity $L_{\rm acc}$ above a few per cent of 
$L_{\rm Edd}$ (see section \ref{eddington}),
the accreting gas tends to radiate
approximately as a blackbody.  In this spectral state one can
theoretically calculate the flux of radiation $F(r)$ emitted by the
accretion disk, and hence obtain the effective temperature profile
$T_{\rm eff}(r) \equiv [F(r)/\sigma]^{1/4}$, where $\sigma$ is the
Stefan-Boltzmann constant.  If the disk emits as a true blackbody at
each radius, it is a simple matter to calculate the total spectral
luminosity $L_\nu d\nu$.  By comparing this quantity with the spectral
flux $F_\nu d\nu$ received at Earth, one obtains an estimate of
$r_{\rm in}^2\cos i/d^2$ (essentially the projected solid angle of the
disk), where $i$ is the inclination angle and $d$ is the distance to
the source.
In a few BH binaries, sufficiently reliable estimates of $i$, $d$ and $M$
are available, and thus an estimate of $r_{\rm in}$ is obtained.  
Identifying $r_{\rm in}$ with
$r_{\rm is}$,  one then obtains $a$. 

A major weakness of this method is that a number of effects will
cause the spectrum of an accretion disk to deviate from a blackbody.
Besides, the method requires accurate estimates of $M$,
$i$, and $d$. 
Therefore, spin estimates obtained by this method should be treated
with caution.  
\vskip \baselineskip
\noindent
{\bf Quasiperiodic Oscillations}
\vskip \baselineskip
For some BH binaries, the power spectrum of intensity variations shows
one or two peaks (more like bumps in some cases) at frequencies of a
few hundred Hz.  The peaks are relatively broad, indicating that they
do not correspond to coherent oscillations but rather to
quasiperiodic oscillations (QPOs). 

One possibility is that the QPO with the
highest frequency in each BH binary corresponds to the circular 
Keplerian frequency of gas blobs at some characteristic radius; it
is plausible that this radius corresponds to the inner edge of the
disk.  Using  equations (\ref{eq931}) and (\ref{eq917}), one can express
the Keplerian frequency of the ISCO as
\begin{equation}
 \Omega_{\rm is}=\frac{1}{M} F(a/M),
 \end{equation} 
where $F(x)$ is a known function of $x=a/M$.
Assuming that $r_{\rm in} =r_{\rm is}$,
one can use this method to estimate $a$ provided an
estimate of $M$ is available.  The method has been applied to a few BH
binaries (for references, see \cite{nar}).  Recently, there has been tentative
evidence for QPOs with a period of 17 minutes in the infrared emission
(Genzel et
al. \cite{gen})
from Sgr A*, the supermassive BH in the Galactic Center .  If the QPOs correspond to the Keplerian frequency at any
radius $r>r_{\rm is}$, then the BH must be rotating with $a>0.5$  \cite{nar}.
A number of QPO frequencies  in the
X-ray flares from Sgr  A* have been identified by Aschenbach et al. \cite{asch}.
Their analysis reveals that the emission from the
inner parts of the accretion disk is quite close to the
BH horizon and they find  $a\simeq 0.99$.
\vskip \baselineskip
\noindent
{\bf Relativistic Iron line}
\vskip \baselineskip
A strong broad spectral line in the
X-ray spectrum of the active galactic nucleus (AGN) MCG-6-30-15
was recently discovered \cite{tana}. They interpreted the line
as fluorescent iron K$\alpha$ emission from cool gas in the accretion
disk.  Similar broad lines were seen in a few other AGN and 
X-ray binaries.
Whereas the rest energy of
the iron line is 6.4 keV, the observed line extends from about 4 to 7
keV. This broadening is due to Doppler blue- and red-shifts as well
as to the gravitational redshift.

The line width and its shape, among other factors, depend on
the radius range over which the emission occurs and
in particular on the position of the innermost radius
of the disk which in turn depends on $a$ (since $r_{\rm in} = r_{\rm is}$). 

Given a system with a broad iron line, and assuming that the radiating
gas follows Keplerian orbits with radii $r \geq r_{\rm is}$, one can
fit the shape of the line profile by adjusting $a$, $i$, and the
emissivity function; the latter is usually modeled as a power law in
radius, $r^{-\beta}$, motivated by the standard disk model
\cite{shak}. 
The effect of $a$ is particularly dramatic.  As the BH spin
increases, the inner edge of the disk comes closer to the horizon
 and the velocity of the gas increases substantially.  This
gives a wider range of Doppler shifts, as well as a larger
gravitational redshift.  The
detection of such extreme levels of broadening may be taken as a strong
indication of a rapidly spinning BH.

In the case of MCG-6-30-15, the data confirm that the emission comes
from a relativistic disk and at least
some of the data sets can be interpreted in terms of a rapidly spinning
BH. Assuming that there is no emission from within the ISCO,
Reynolds et al. \cite{rey} estimate $a>0.93$.  Among BH binaries, the
source GX 339-4 shows a broad iron line which seems to indicate
$a>0.8$ (Miller et al. \cite{mil}). 

In spite of some weaknesses, the method has the advantage that it
requires no knowledge of the BH mass or of the distance, and it
solves for the disk inclination $i$ using the same line data from
which $a$ is estimated.  

The variability of the line with time
means that it will be challenging to make fundamental tests of gravity
with this method.  On the other hand, the variability could provide
interesting opportunities to study disk dynamics and turbulence
\cite{armit,rey2}
(movies courtesy P. Armitage).

\section{Supermassive Black Holes}
\label{supermass}
It is now widely accepted that quasars and active galactic nuclei
are powered by accretion onto massive black holes
\cite{lynd,Madejski,deZeeuw}. 
Further, over the last few years there
has been increasing evidence that massive dark objects may reside
at the centers of most, if not all, galaxies
\cite{rees,melia}. In several cases, the best
explanation for the nature of these objects is that they are
supermassive black holes, with masses ranging from $10^6$ to
$10^{10}$ solar masses. 
Comprehensive lists of about 30 supermassive BHs 
at galactic centers may be found
in~\cite{fer1,MerFer,KorGeb,korm}.  

\subsection{Masses and Radii}    
  The main
criterion for finding candidates for such black holes is the
presence of a large mass within a small region.
The mass and the size  are estimated  using mostly the following three
methods:
 gas spectroscopy, maser interferometry, and measuring the
motion of stars orbiting around the galactic nucleus.

\subsubsection{Gas Spectroscopy}
An example of measurements via gas spectrography is
given by the analysis of the
Hubble Space Telescope  observations of the radio galaxy
M~87~\cite{mac,tsv}. 

A spectral analysis
shows the presence of a disklike structure of ionized gas in the
innermost few arc seconds in the vicinity of the nucleus of M~87.
The velocity of the gas measured by spectroscopy
 at a distance from the nucleus of the order
 20 pc $\approx 6 \times 10^{14}$~km, shows that the gas
 recedes from us on
one side, and approaches us on the other, with a velocity
difference of about $ 920$~km~s$^{-1}$ . This leads to a mass of
the central object of $\sim 3 \times 10^{9}$~$M_{\odot}$, and no
form of matter can occupy such a small region except for a black
hole. 
This is the most massive black hole ever observed.
\subsubsection{Maser Interferometry}
A clear and compelling evidence for 
black holes has recently been discovered  
in the radio regime: 
H$_2$O masers orbiting compact supermassive
central objects. 
The structure of accreting material around 
the nearby galaxy NGC 4258 ($d\simeq 6.4$ Mpc) has been studied in detail \cite{her,mor,pie}. 
with the aid of very long baseline
interferometry (VLBI), which provides an angular resolution as fine as 
200~$\mu$as
(microarcseconds)
at a wavelength of 1.3~cm
and a spectral resolution of 0.1 km$\,$s$^{-1}$ or less,
radio interferometry measurements have shown
that the gas follows circular orbits with a nearly perfect Keplerian
velocity profile ($v \propto r^{-1/2}$,  see Fig. 1 in \cite{nar}). 
 Furthermore, the
acceleration of the gas has been measured and it too is consistent
with Keplerian dynamics (Bragg et al. \cite{bra}).  From these measurements
it is inferred that there is a dark object with a mass of
$3.5\times10^7 M_\odot$ confined within $\sim$~0.13 pc $=4\times10^{12}$ km of the
center of NGC 4258.  The case for this dark mass being a BH is quite 
strong.
\subsubsection{Virial Mass}
A simple method   
to obtain a mass estimate is
based on the virial theorem.
It uses the measured velocity
dispersion of stars in the central region
(ref. \cite{carro}, p. 1007).
For simplicity, we restrict attention to a spherical cluster of radius $R$
with $N$ stars, each of mass $m$, so the total mass of the bulge
is $M=Nm$.

The time-averaged kinetic and potential energies 
of stars in the galaxy's central region
are related by the equation 
(ref. \cite{carro}, p. 54-56)
\begin{equation}
\frac{1}{2}\left< \frac{d^2I}{dt^2}\right>- 2\left<K\right>=
\left<U\right>\, ,
\label{eq110}
\end{equation}
where $I$ is the region's moment of inertia. If the galaxy is in equilibrium,
then $\left< d^2I/dt^2\right>=0$, resulting in the usual statement of the virial 
theorem
\begin{equation}
- 2\left<K\right>=
\left<U\right> \, .
\label{eq111}
\end{equation}
Furthermore, for a large number of stars, the central bulge will look
the same (in statistical sense) at any time, and the time averaging can be dropped.
So for $N$ stars of equal mass, we find
\begin{equation}
- \frac{m}{N}\sum v_i^2=\frac{U}{N} \, .
\label{eq112}
\end{equation}
The average velocity squared is
\begin{equation}
- \frac{1}{N}\sum v_i^2=
\left<v^2\right>=
\left<v_r^2\right>+
\left<v_\theta^2\right>+
\left<v_\phi^2\right>=3\left<v_r^2\right>=3\sigma_r^2 \ ,
\label{eq113}
\end{equation}
where $\sigma_r$ is the dispersion in the radial velocity.

\begin{description}
\item[Velocity dispersion] -----------------------------------------------------------------------------------------
\\ For a region in space with a large number of stars $N$,
 $\sigma_x$ 
measures the spread in the $x$-component of the peculiar velocities of stars
and 
is defined as
\begin{equation}
\sigma_x=\frac{1}{N}\left[\sum_i v_{ix}^2 \right]^{1/2} .
\label{eq114}
\end{equation}
It is equal to
the standard deviation of the velocity distribution
in the special case when $\left<v_x\right>=0$.
\\
----------------------------------------------------------------------------------------------------------------
\end{description}

\noindent
Using the (approximate) gravitational
 potential energy of a spherical distribution of the total mass
 (exercise)
\begin{equation}
U=-\frac{3}{5}\frac{GM^2}{R} \, ,
\label{eq115}
\end{equation}
equations (\ref{eq111}) and (\ref{eq112}) yield
\begin{equation}
M_{\rm virial}=\frac{5R\sigma_r^2}{G} \, ,
\label{eq116}
\end{equation}
where the mass obtained in this way is called the {\em virial mass}.

This equation can be used to estimate a virial mass for the central BH 
of M31 (Andromeda). The central radial velocity dispersion is 
measured to be approximately 240~km/s within 
0.2 as (arcseconds). Given the distance to Andromeda 
of 770~kpc, 0.2 as corresponds to 
$R \simeq 0.8$~pc. This gives a total mass of roughly
\begin{equation}
M_{\rm virial}= 6\times 10^7 M_\odot \, ,
\label{eq117}
\end{equation}
 within a sphere of radius 0.8~pc. 

This is, of course, just an order of magnitude estimate.
Other estimates  for the mass of the supermassive BH
of M31
range from about $10^6$ to $10^7 M_\odot$.

\subsection{Sagittarius A$^*$}
Perhaps the most convincing evidence for a black hole comes from the
center of our own galaxy that coincides with 
the enigmatic strong radiosource Sgr A$^{*}$.
The existence of a dark massive object at the center of the Galaxy
has been inferred from 
the motions of stars and gas in its vicinity. 
The motions of the stars were observed and recorded for many years by
two independent groups 
\cite{schod1,eis,eck2,ghez3,ghez4}.
High-resolution infrared observations made it possible
to follow the orbits of individual stars around this object (Sch\"odel
et al. \cite{schod1,eis}; Ghez et al. \cite{ghez3,ghez4}).
The movies (1st movie courtesy R. Genzel;
2nd movie  
courtesy A. Ghez \cite{ghez5}) 
show time-elapsed images of the Galactic Center region
revealing the (eccentric) orbits of several stars.

The projected positions of the star S2(S0-2)
that was observed during the last decade \cite{schod1,eis}  
suggest that
S2(S0-2) is moving on a
Keplerian orbit with a period $P$ of 15.2 yr around
Sgr A$^{*}$ and the estimated semimajor axis $a$ 
of the order of 4.62 mpc.
 
Then, neglecting the star mass, Kepler's third law (\ref{eq90})
gives
\begin{equation}
M_{\rm SgrA*}= \frac{4\pi^2 a^3}{G P^2}= 3.7\times 10^6 M_\odot \, .
\label{eq118}
\end{equation}

The salient feature of the new adaptive optics
data is that, between April and May 2002,
S2(S0-2) apparently sped past the point of
closest approach 
 with a velocity $v$ $\sim$
6000 km/s at a distance of about 17 light-hours
\cite{schod1} or 123 AU from Sgr A$^{*}$.
This implies that an enormous mass of the central object is 
concentrated in a very small volume,
strongly
suggesting that the object is a BH.

Another star, S0-16 (S14), which was observed
during the last few years by Ghez et al.
\cite{ghez4}
with the Keck telescope in Hawaii, recently made
 a spectacular U-turn, crossing the
point of closest approach at an even smaller
distance of 8.32 light-hours or 60 AU from
Sgr A$^{*}$ with a velocity $v$ $\sim$
9000 km/s. Ghez et al. thus conclude
that the gravitational potential around
Sgr A$^{*}$ has an approximately $r^{-1}$ form,
for radii larger than 60 AU, corresponding to
1169  
$R_{\rm SCH}$, where  $R_{\rm SCH}= 2M =$ 
 0.051~AU for $M= 2.6
\times 10^{6} M_{\odot}$.

\subsection{Supermassive BHs in Active Galactic Nuclei}
Many galaxies possess extremely luminous
central regions, 
with luminosity (in particular the luminosity of X-ray radiation)
exceeding the luminosity of ordinary galaxies
by several orders of magnitude.
These luminous central regions are called {\em active galactic nuclei} (AGN).
To this class belong the so-called {\em quasistellar objects} (QSO)
and {\em quasars} \footnote{Quasar is short for {\em quasistellar radio source}.
Quasars are radio-loud whereas  QSOs are radio-quiet . These names are sometimes
confused in the literature. Both terms, {\em QSO} and {\em quasar},
are often used to refer to both types of objects \cite{carro}.}.

What makes these galaxies ``active'' is the emission of enormous amounts of energy from their 
nuclei. Moreover, the luminosities of active galactic nuclei fluctuate on very short
 time scales -- 
within days or sometimes even minutes. 
The time variation sets an upper limit to the size of the 
emitting region. For this reason, we know that the emitting regions of active 
galactic nuclei are only 
light-minutes or light-days across, 
making them less than one ten-millionth the size of the galaxy in 
which they sit.

How could a luminosity hundreds of times that 
of an entire galaxy be emitted from a volume billions of times smaller? 
Of all proposed 
explanations, only one has survived close scrutiny: 
the release of gravitational energy by matter falling 
towards a black hole \cite{lynd,fer}.
Even using an energy source as efficient as gravity, the black holes
in active galactic nuclei would need to be supermassive in order
to produce the luminosities of quasars. 

\subsubsection{Eddington Limit}
\label{eddington}
The most efficient way of generating energy is by the release
of gravitational potential energy through mass accretion.
For example, a simple calculation shows  that for matter falling straight down
onto the surface of a 1.4~$M_\odot$ neutron star, about 21\% of the rest
mass is released. This is almost 30 times larger than the energy that hydrogen
fusion can provide.

The efficiency may be even larger if the matter is
accreted through an accretion disk of a rotating black hole.
The {\em accretion luminosity} (i.e., the radiation energy released per unit of time)
generated by a mass accretion rate, $\dot{M}$, through the disk may be written as
 \begin{equation}
L_{\rm disk}= \eta \dot{M},
\label{eq119}
\end{equation}
where $\eta$ is the radiative efficiency of the disc
equal to the binding energy of the innermost stable circular orbit
per unit rest mass. 
As we have shown in section \ref{innermost}, $0.0572<\eta<0.423$,
with the lower and upper  bounds  for a nonrotating and for a maximally
rotating BH, respectively.

However, the radiation is interacting with the accreting gas
and there is a limit to the luminosity above which the radiation pressure,
acting against gravitational attraction, exceeds gravity
and thereby stops the accretion. 

Consider a fully ionized hydrogen plasma accreting near the surface of a compact
object of mass $M$. The upward force on the infalling matter is mainly
due to the interaction of radiating photons with electrons in the plasma.
If the photon luminosity is $L$,
the number of photons  $N_\gamma$ crossing unit area per unit time 
at radius $r$ is
\begin{equation}
N_\gamma= \frac{L}{\epsilon_r 4\pi r^2}\, ,
\label{eq120}
\end{equation}
where $\epsilon_r$ is the mean energy transferred radially per collision.
The number of collisions per electron per unit time is $\sigma N_\gamma$, where 
$\sigma$ is the photon-electron scattering cross section. The force per electron
is just the rate at which the momentum is deposited radially per unit time, so we 
multiply by $p_r=\epsilon_r$ to obtain
\begin{equation}
F=\sigma N_\gamma p_r= \frac{L \sigma}{4\pi r^2}\, .
\label{eq121}
\end{equation}
In order for accretion to occur, the gravitational force
per electron (acting via the proton) 
\begin{equation}
F_{\rm grav}= -\frac{{\cal M} m_p}{r^2}
\label{eq122}
\end{equation} 
must exceed the radiation force (\ref{eq121}).
Here, ${\cal M}\simeq M$ is the enclosed mass at  radius $r$.
Equating the radiation force (\ref{eq121}) with the 
gravitational force (\ref{eq122}) sets an upper bound
to accretion luminosity known as the {\em Eddington limit}
\begin{equation}
L_{\rm Edd}= 4\pi \frac{{\cal M} m_p}{\sigma}\, .
\label{eq123}
\end{equation} 
The dominant  electron-photon process in a highly
ionized hydrogen gas is the scattering of photons 
by free electrons ({\em Thomson scattering}) with the cross section
\begin{equation}
\sigma_{\rm T}=\frac{8\pi}{3}\left(\frac{e^2}{m_e}\right)^2=6.65\times 
10^{-25}\: {\rm cm}^2 .
\label{eq124}
\end{equation} 
This gives
\begin{equation}
L_{\rm Edd}= 1.3 \times 10^{38}\left(\frac{M}{M_\odot}\right)
\:\mbox{erg s}^{-1} .
\label{eq125}
\end{equation} 
\subsubsection*{Relativistic Eddington Limit}
In a strong gravitational field the expression 
(\ref{eq122}) for the gravitational force is no longer valid.
Here we derive the general relativistic expression 
for the force per electron (acting via the proton)
under the assumption that the energy density of the fluid
is still nonrelativistic, i.e., $\rho=m_p n$ and 
$p\ll\rho$.

Consider a spherical shell of radius $r$
with thickness $dr$. The force acting on the shell element of surface area $dS$  is
\begin{equation}
dF =  dS\, dp =  d{\cal V} (1-2{\cal M}/r)^{1/2}\frac{dp}{dr} \, ,
\label{eq400}
\end{equation}
where 
\begin{equation}
d{\cal V}=dr\,dS (1-2{\cal M}/r)^{-1/2}
\end{equation}
is the proper volume  element 
and $dp$ is the difference between radial pressures at $r$ and $r+dr$.
Next, we substitute $dp/dr$ by the right-hand side of the TOV equation 
(\ref{eq48}) in which we neglect the pressure 
(since $p\ll \rho$ by assumption).
According to (\ref{eq45}), the number of particles in the shell element is
$d{\cal N}= n d{\cal V}$.
Then, the gravitational force per particle is 
\begin{equation}
F_{\rm grav}=\frac{dF}{d{\cal N}}=
-\frac{m_p {\cal M}}{r^2}(1-2{\cal M}/r)^{-1/2} .
\label{eq401}
\end{equation}
 Equating $|F_{\rm grav}|$
thus obtained   with  the photon force
(\ref{eq121}),
we find 
\begin{equation}
L_{\rm Edd}
=\frac{4\pi m_p{\cal M}}{\sigma}(1-2{\cal M}/r)^{-1/2} .
\label{eq404}
\end{equation}
Hence,  compared with (\ref{eq123}), the Eddington limit  in the strong gravitational field 
is larger by
a relativistic correction factor.

\subsubsection{Radius and Mass Estimate}
\label{agn-mass}
The luminosity of a typical quasar
varies in time with a typical period  of 1 hr.
This sets the  upper limit to the radius of
about
\begin{equation}
R\simeq 7\:{\rm AU}=1.1\times 10^{9}\:{\rm km}.
\label{eq126}
\end{equation} 
Considering that AGNs are the most luminous objects known, this is
an incredibly small size.

The typical quasar luminosity of 5 $\times 10^{46}$ erg s$^{-1}$
is equivalent to more than 500 galaxies of the Milky Way size! 
Now, the constraint
that the luminosity must be less than the Eddington limit,
$L< L_{\rm Edd}$,  provides a lower limit for the mass of the central
object
\begin{equation}
M>M_0 =3.3 \times 10^8 M_\odot\, .
\label{eq127}
\end{equation}
The mass limit $M_0$ is quite close to the mass 
$M_{\rm BH}=R/(2G)= 3.7 \times 10^8 M_\odot$ of a BH  the Schwarzschild radius of which
is equal to the radius  of the quasar $R$ estimated above.
This fact supports the idea that supermassive BHs are responsible for
powering AGNs.
(Carroll and Ostlie \cite{carro}, p. 1181)

The estimated value of the
lower mass limit is somewhat smaller if the general relativistic
correction to the Eddington limit is taken into account.
Using the relativistic expression (\ref{eq404}), we find
\begin{equation}
M>M_0\left(\sqrt{1+\frac{M_0^2}{4M_{\rm BH}^2}}-\frac{M_0}{2M_{\rm BH}}\right)
=2.14 \times 10^8 M_\odot\, .
\label{eq405}
\end{equation}

\subsubsection{Estimate of the AGN Efficiency}

According to equation  (\ref{eq119}), 
the radiative efficiency $\eta$ of an accretion disk is 
defined as the energy
it radiates per unit accreted mass.
As shown in section \ref{innermost}, $\eta$  equals 
the binding energy of gas at the ISCO, which in turn
depends on the BH spin parameter $a$.   

In a typical accretion system, one can easily measure 
the accretion luminosity $L_{\rm disc}$
(provided the distance is known), but one practically never has an
accurate estimate of the mass accretion rate $\dot M$,
 so one cannot calculate $\eta$
for an individual AGN with the
precision needed to estimate $a$.  However, a rough estimate can be 
made for an average AGN.
From observations of high redshift AGN, one
can estimate the mean energy radiated by supermassive BHs per unit
volume of the universe.  Similarly, by taking a sample of supermassive
BHs in nearby galaxies, one can estimate the mean mass in BHs per unit
volume of the universe at present.  Assuming that supermassive BHs
acquire most of their mass via accretion (a not unreasonable
hypothesis), one can divide the two quantities to obtain the mean
radiative efficiency of AGN.  The current
data suggest an efficiency $\eta\sim 0.1-0.15$ for supermassive BHs on
average (Elvis et al \cite{elv}, Yu and Tremaine \cite{yu}).
Such large values of $\eta$ are possible only if supermassive
BHs have significant rotation.

It should be noted that this is only a statistical
result for the population of AGN as a whole, and the method does not say
anything about the rotation of any specific BH.

\section{Intermediate Mass BHs}
So far we have been concerned with either the stellar mass BHs 
($M \sim 3.5 -20 M_{\odot}$) or the supermassive BHs 
($M > 10^6 M_{\odot}$). Are there BHs in the intermediate
mass range $10^2-10^4 M_{\odot}$? There is {\em a priori}
no reason for such BHs not to exist.
 
There is a tentative evidence based on the Eddington limit
that the  intermediate mass BHs do exist.
 In several nearby galaxies 
a number of ultraluminous X-ray sources have been detected
\cite{mush}. As their luminosities 
of the order of $10^{40}$ or more exceed the Eddington limit 
(\ref{eq125}) of
 10 $M_{\odot}$ BHs, these objects are argued to be  good candidates
 for the intermediate mass BHs \cite{colb}.
 However, large apparent luminosity may, under special conditions,
 be produced by stellar mass BHs,
 e.g, by a  
 supercritical accretion
 (for details and references, see Poutanen et al.\ \cite{pou}).
 Hence, it is at present unclear what exactly the ultraluminous
 X-ray sources are. Dynamical mass measurements would settle the issue
 but, unfortunately, none of the sources has a binary companion
 to provide a robust mass estimate.

\section{Observational Evidence for the Horizon}
\label{evidence}
To prove that a BH-like object is indeed a BH, one needs to demonstrate that it
possesses an event horizon. The major tests for a BH horizon are based on
i) advection-dominated accretion flows (ADAF), ii) X-ray bursts, and iii)
direct imaging \cite{nar,abr,hug}
\subsection{ADAF}
Advection dominated accretion flows 
\cite{nar1,nar5}
describe accretion with very low radiative efficiency in which the
energy released by viscosity friction removing the angular momentum from the
accreting matter is not radiated away but stored in the flow. If an ADAF
forms around a BH, the stored energy will be lost forever under
the event horizon, whereas if the accreting object is a NS, this
energy must be radiated away once matter lands on its surface.
Therefore, the argument goes, BHs should be dimmer than NSs
if an ADAF is present in both cases.
 
 Narayan, Garcia, and McClintock \cite{nar2}
 suggested that
 precisely such a comparison could be done using X-ray binaries.
 Most of the known stellar-mass BH candidates are
in a class of X-ray binaries called X-ray novae.  These systems are
characterized by a variable mass accretion rate, and tend to spend
most of their time in a quiescent state with a very low accretion rate $\dot M$
 and accretion luminosity
$L_{\rm acc}$.  Only occasionally do they go into outburst, when they
accrete with high $\dot M$ and become bright.  Spectral observations
of quiescent BH binaries can be explained in terms of an ADAF.
Narayan et al. \cite{nar2} compared  quiescent
luminosities of X-ray  binaries supposed to contain BHs with those of
neutron-star X-ray  binaries and realized that, in accordance with the prediction
of the ADAF model, systems containing black-hole ``candidates"
are dimmer. They came to the conclusion that they found evidence for the
presence of event horizons. 
However, this conclusion has been challenged (for details and references, see
\cite{nar,abr}).
\subsection{X-ray Bursts}
In some X-ray binaries, 
X-ray bursts are observed in addition to the quiescent X-ray luminosity. 
In a typical X-ray burst luminosity increases 
up to nearly the Eddington limit in less than a second, and the flux then declines over a period of a few seconds or a few tens of seconds.
 Remarkably, no X-ray bursts of this kind were observed in any BH binary.
 Narayan and Heyl \cite{nar3} argued that the lack of bursts is a strong
 evidence for the horizon in BH candidates.
 
 The explanation  for such an absence of X-ray bursts is quite simple.
 At present, it is widely accepted that X-ray bursts 
 arise
from thermonuclear detonation of accreted material
\cite{nar3,nar4}.  Accreted material builds up on the
surface of the compact object and is compressed by the object's
gravity.  After sufficient material accumulates, it undergoes unstable
thermonuclear burning, which we observe as an X-ray burst.

A key point is that the compact object {\it must have
a surface}.  Material cannot accumulate on an event horizon, and so no
bursts can come from an X-ray binary whose compact object is a BH.
For more details, references, and critical discussion, see \cite{nar, abr}

\subsection{Direct Imaging}

The most promising line of search for a direct evidence 
is to construct an image of the region near the event horizon
using interferometry.

For definiteness, consider  a nonrotating BH with a horizon
at radius $R$.  Because of strong gravitational lensing
 in the vicinity of
the BH, a distant observer will see an apparent boundary of the BH at
a radius of $3\sqrt{3}R/2$ \cite{fal}.  Rays with
impact parameters inside this boundary intersect the horizon, while
rays outside the boundary miss the horizon.  The angular size of the
boundary, e.g., for   
   Sgr A$^*$ with $M\sim
4\times10^{6}M_\odot$ and at a distance of 8 kpc, is
 $\sim 0.02$ mas which is not beyond reach.  The
supermassive BH in the nucleus of the nearby (15 Mpc) giant elliptical
galaxy M87, with a mass of $3\times10^9M_\odot$ and an expected
angular size of $\sim 0.01$ mas, is another object of interest
\cite{kri}.

The best angular resolution achievable today is with radio
interferometry, where angles less than 1 mas are routinely resolved.
In the not too distant future, it should be possible to operate
interferometers at wavelengths $\lambda < 1$ mm and with baselines as
large as the diameter of the Earth $b\sim10^4$ km. For details, see
Falcke et al. \cite{fal}.

\section{Alternatives to Supermassive BHs}
\label{alternatives}
 Given the accumulated evidence for supermassive compact objects  ranging from
a few $10^{6}M_{\odot}$ to a few $10^{9} \, M_{\odot}$,
the existence of black-hole-like objects is beyond doubt \cite{melia,korm}.
What still remains an issue is whether these supermassive objects are BHs
with the Schwarzschild (or Kerr) metric describing
the physics of the interior, or some other objects built out of
more or less exotic substance but with a regular behavior in the interior.
A  standard astrophysical  scenario
in the form of a compact cluster of dark stars (e.g., neutron stars or brown dwarfs)
 although not entirely excluded, is quite unlikely.
 It has been demonstrated that, in the case of NGC 4258 and our Galaxy, 
  such a cluster 
  would be short-lived and would either ``evaporate" or
become a BH in much less time than the age of the Galaxy 
\cite{maoz}.

A number of alternatives to classical BHs have been proposed
with no singularities in the interior.
The three representative  models are: 1) Neutrino (or neutralino) stars,
2) Boson stars, 3) Dark energy stars 

\subsection{Neutrino Stars}
\label{neutrino}

Here we use the term {\em neutrino stars} as a generic name
for any degenerate fermion star composed of neutral weakly interacting 
fermions, e.g., neutrinos or supersymmetric fermionic partners such as neutralinos, gravitinos, and axinos. 
The simplest model proposed  for supermassive compact objects at the
galactic centers is a self-gravitating
degenerate fermion gas composed of, e.g.,
 heavy sterile neutrinos
 \cite{viol5,viol6,bil7,bil8,mun9}. 
 Sterile neutrinos in the keV mass range have recently been
 extensively discussed as dark-matter candidates \cite{aba,kus}

  As we have seen in the example
 of a neutron star,  a self-gravitating ball of degenerate
fermionic matter is supported against  gravitational collapse
 by the degeneracy
pressure of fermions due to the
Pauli exclusion principle, provided the total mass is below
the OV limit (\ref{eq900}), with an $m^{-2}$ functional dependence
on the fermion mass $m$.

Let us assume that the most massive objects
are sterile neutrino stars
 at the OV limit with
$M_{\rm{OV}}\simeq 3 \times 10^{9}$~$M_{\odot}$,
such as the supermassive compact dark object
at the center of the radio-galaxy M87
\cite{mac}.
 Using  (\ref{eq900})  we find that the neutrino mass
 required for this scenario is
\begin{eqnarray}
m
   &\simeq&
 14\:{\rm keV}
 \;\;\;\;\; {\rm for}
\;      g=2,
\nonumber \\
 m
   &\simeq&
 12\:{\rm keV}
 \;\;\;\;\; {\rm for}
\;      g=4.
\label{eq1100}
\end{eqnarray}
From (\ref{eq900}) and (\ref{eq901}) it follows that
a neutrino star of mass
$M_{\rm{OV}}=3\times 10^{9}~M_{\odot}$
would have a radius
$R_{\rm{OV}}=4.4466 R_{\rm Sch}=3.9396\times 10^{10}$ km
$= 1.52$ light-days,
 where $R_{\rm Sch}=2G M_{\rm{OV}}$ is the
Schwarzschild radius for $M_{\rm{OV}}$.
Thus, at a distance of a few Schwarzschild radii away from the
 supermassive object,
there is little difference between a neutrino star
at the OV limit
and a BH, in particular
since the last stable orbit around a BH
already has a radius of $3~R_{\rm Sch}$.

Now, we wish to investigate the possibility that this scenario 
can be extrapolated to explain all observed supermassive BH,
and in particular  Sgr~A$^*$, the supermassive
compact dark object at the Galactic center.
As the mass of Sgr~A$^*$ is by about a factor
of $10^3$ lower than the OV limit above, we can use the
Newtonian mass-radius scaling relation (\ref{eq64})
derived from
 the Lane-Emden equation (\ref{eq59})
with the polytropic index $n = 3/2$. We find
\begin{equation}
R  = 4.51 \,
\frac{m_{\rm Pl}^2}{m^{8/3}M^{1/3}} \, \left( \frac{2}{g}
\right)^{2/3}  
   =  7.1406 \times 10^5 
  \left( \frac{12 \: \mbox{keV}}{m}
  \right)^{8/3}  \left( \frac{4}{g} \right)^{2/3} 
  \left( \frac{M_{\odot}}{M} \right)^{1/3} \: \mbox{AU}.
  \label{eq1000}
\end{equation}
Here, 1~mpc = 206.265 AU. The degeneracy factor
$g$ = 2 describes either spin 1/2
fermions (without antifermions) or spin
1/2 Majorana fermions. For Dirac fermions 
(or spin 3/2 Majorana fermions),
 we have $g$ = 4.
Using the canonical value
$M = 2.6 \times 10^{6} M_{\odot}$,
we find $R\simeq 5200 AU$
to be compared with the observational upper limit \cite{ghez4}
 $R$ $\leq$ 60 AU for Sgr~A$^*$.
 Hence, a neutrino star made of $m=12-15$ keV neutrinos is 
 ruled out as an explanation for Sgr~A$^*$.

To be able to explain Sgr~A$^*$ as a neutrino star,  
we need  a minimal fermion mass 
 $m_{\rm min}$ = 63.9 keV/$c^{2}$
for $g$ = 4. 
 The maximal mass of a neutrino star made of these fermions,
 as given by the OV 
 limit (\ref{eq900}), is
 \cite{bil1}
\begin{equation}
M_{\rm OV}^{\rm max} = 1.083 \times 10^{8} M_{\odot} \; \; .
\end{equation}

Clearly, a neutrino star scenario which would cover the whole
mass-radius  range 
of compact supermassive galactic centers is ruled out.
However, 
a ``hybrid'' scenario \cite{bil1} 
in which
 all supermassive compact
dark objects with masses $M > M_{\rm OV}^{\rm max}$
are black holes, while those with
$M \leq M_{\rm OV}^{\rm max}$ are neutrino stars,
is not excluded.

At first sight, such a hybrid scenario does not
seem to be particularly appealing. However, it
is important to note that a similar scenario is
actually realized in nature, with the co-existence
of neutron stars with masses
$M \simlt 3 M_\odot$, and stellar-mass BHs
 with mass $M \simgt 3 M_\odot$, as observed
in  binary systems in the Galaxy.

An indirect support for this scenario is the  absence
of clear evidence for intermediate mass BH candidates, which is difficult to
explain in the conventional BH scenario
(in which BHs are all baryonic).
If the hybrid scenario were realized in nature, 
the intermediate mass neutrino stars
would exist but having very large radii 
would be rather dilute and hence difficult to detect. 

\subsection{Boson Stars}

Boson stars are static configurations of self-gravitating 
(with or without self-interaction) complex scalar fields.
In self-interacting scalar field theories, such as $\phi^4$ theory,
there are cases 
where a homogeneous condensate is a stable ground state,
known as the Bose-Einstein (BE) condensate.
Hence, a boson star being  basically a self-gravitating
BE condensate may be regarded as a $T\rightarrow 0$ 
limit of a self-gravitating boson gas \cite{bil4}.
     The ground state of a
     condensed cloud of charged bosons of mass $m$,
     interacting only gravitationally
     and having a total
 mass $M$ below a certain limit
 of the order $M_{\rm Pl}^2/m$,
 is a stable spherically symmetric configuration
 \cite{kau}
which is usually referred to as
 mini-soliton star \cite{fri2}
 or mini-boson star \cite{jet}.
 For a recent review and a comprehensive list of references,
 see Schunck and Mielke \cite{schu}.

 The gravitational collapse of a self-interacting BE
 condensate is prevented by a repulsive 
  self-interaction, e.g., in the form of
$\lambda |\Phi|^4$.
That makes it 
 astrophysically interesting as 
its maximal mass is $\sim m_{\rm Pl}^3/m_B^2$,  hence 
comparable with the mass of a neutron star or a neutrino star.

Seidel and Suen
have demonstrated that boson stars
may be formed through a dissipationless mechanism,
called gravitational cooling
\cite{sei}.
Boson stars have recently attracted some attention
as they may well be candidates for
nonbaryonic dark matter
\cite{mie}.
 
The action of the gravitationally coupled complex scalar 
field $\Phi$ reads
\begin{equation}
S_{\rm BS} =\int d^4x\, \sqrt{- \det g} \left(-\frac{R}{16\pi}
 +  {\cal L}_{\rm KG} \right),
 \label{lagrange}
\end{equation}
with the Klein-Gordon Lagrangian 
\begin{equation}
{\cal L}_{\rm KG} = 
   g^{\mu \nu} (\partial_\mu \Phi^*) (\partial_\nu \Phi )
             - m^2 |\Phi|^2 -U(|\Phi|^2)  \, .
\end{equation}
By varying the action with respect 
to $\Phi$ one obtains the
 Klein--Gordon equation
\begin{equation}
      \left (\mbox{\Large $\Box$} + \frac {dU}{d| \Phi |^2} \right ) \Phi
             =  0 \, ,
\label{eq159}
\end{equation}
and the variation with respect to $g_{\mu\nu}$ yields
 Einstein's equations (\ref{eq201}) 
in which
\begin{equation}
T_{\mu \nu }
 =  (\partial_\mu \Phi^\ast )(\partial_\nu \Phi )
  + (\partial_\mu \Phi )(\partial_\nu \Phi^\ast )
  - g_{\mu \nu } {\cal L}_{\rm KG} , . 
\end{equation}
Although this energy-momentum tensor is not in the perfect fluid form
(\ref{eq001}, we can still identify the radial pressure $p$ and
the density $\rho$
with $-T^r_r$ and $T^0_0$, respectively.

Given the  interaction potential $U$,
one has to solve a coupled system of Klein-Gordon and Einstein 
equations in spherically symmetric 
static spacetime. Unlike in the case of a fermion star,
where the information about matter was provided by
 the equation of state, the matter here is described by the KG equation
 (\ref{eq159}).
The stationarity ansatz
\begin{equation}
\Phi (r,t)=\frac{1}{\sqrt{2}}\varphi(r) e^{-i\omega t} 
\end{equation}
 describes a spherically symmetric bound 
state of the scalar field, where the frequency 
$\omega $ is determined by the asymptotic condition 
$\varphi(r)\rightarrow 0$ as $r\rightarrow \infty$.

It turns out that even the simplest case $U=0$ has a nontrivial
ground-state solution, called a {\em mini-boson star}.
The gravitational collapse of a mini-boson star is  prevented by
Heisenberg's uncertainty principle.
This provides us also with  crude mass estimates: For a boson to be confined 
within the star of  radius $R_0$, the Compton wavelength has to satisfy
$\lambda_\Phi= 2\pi/m \simeq 2R_0$. On the other hand,
the star's radius    
should be larger than the Schwarzschild radius, 
$R_0 > R_{\rm Sch}=2M/m_{\rm Pl}^2$  
 in order to avoid instability 
against complete gravitational  collapse.
In this way  we obtain an estimate
\begin{equation} 
M < \frac{\pi}{2}m_{\rm Pl}^2/m \, .
\end{equation}
This upper bound is slightly larger than 
 the so-called  {\em Kaup limit},
\begin{equation}
M_{\rm max}= 0.633 m_{\rm Pl}^2/m =
8.4639 \times 10^{-10} \left(\frac{1\: {\rm eV}}{m}\right)\: M_\odot
\label{eq128}
\end{equation}
 obtained numerically~\cite{jet}.
 Clearly, mini-boson stars are irrelevant
to astrophysical context, unless the boson mass $m$
is ridiculously small.

This result was later extended by Colpi et al.~\cite{colpi} for 
interacting bosons with  a  quartic self-interaction
\begin{equation}
U =
\lambda|\Phi|^4 \,.
\end{equation}
 In this case, the maximal mass   is obtained as \cite{colpi,mie2}
\begin{equation}
M_{\rm max}= \frac{1}{\sqrt{8\pi^3}}\sqrt{\lambda} \frac{m_{\rm Pl}^3}{m^2}=
0.1\sqrt{\lambda} \left(\frac{1\:{\rm GeV}}{m}\right)^2 M_\odot
\label{eq129}
\end{equation}
quite similar to the fermion star mass limit.
The effective radius of an interacting boson star in the Newtonian
regime turns out to be independent of central density 
and in a good approximation is equal to
\begin{equation}
R_{\rm eff}= \frac{\pi}{\sqrt{8\pi}}\sqrt{\lambda} \frac{m_{\rm Pl}}{m^2}=
1.5096 \sqrt{\lambda} \left(\frac{1\:{\rm GeV}}{m}\right)^2 \:
{\rm km} .
\label{eq131}
\end{equation}
Hence, boson stars with a quartic self-interaction all have the same radius.
This behavior is quite different from the fermion star case, where the radius
 scales with the star mass as $R \sim M^{1/3}$ 
 (see equation (\ref{eq1000})).
 
 The results are summarized in the table
 \vskip .2in
\begin{tabular}{|l|c|c|} \hline
              & Maximum       & Effective     \\
 Object       &    Mass   &   Radius         \\  \hline
 Fermion Star  & $M_{\rm OV}=0.384 m_{\rm PL}^3/m^2$
 & $R_{\rm OV}= 3.36 m_{\rm Pl}/m^2$   \\
Mini-Boson Star &$M_{\rm Kaup}=0.633 m_{\rm PL}^2/m$
&$ R \sim 1/m$  \\  
Boson Star &$M_{\rm max}=1/(\sqrt{8\pi^3})\sqrt{\lambda} m_{\rm Pl}^3/m^2$
& $R_{\rm eff}= \pi/(\sqrt{8\pi})\sqrt{\lambda} m_{\rm Pl}/m^2$      \\  \hline
\end{tabular}
\vskip .2in
 
Could boson stars  fit the whole range of compact supermassive galactic centers?
To answer this question, we proceed as in the case of neutrino stars. 
We assume that the most massive objects
are boson stars with maximal mass, i.e.,
$M_{\rm{max}}\simeq 3 \times 10^{9}$~$M_{\odot}$,
for the supermassive compact dark object
at the center M87.
From this we find
\begin{equation}
\sqrt{\lambda} \left(\frac{1\:{\rm GeV}}{m}\right)^2 = 3\times 10^{10} ,
\label{eq132}
\end{equation}
which also fixes the radius given by (\ref{eq131})
\begin{equation}
R=4.53 \times 10^{10} \: {\rm km} = 303 \: {\rm AU} \, ,
\label{eq133}
\end{equation}
quite close to the radius $R_{\rm OV}$ of the
neutrino star  discussed in section \ref{neutrino}.
Since the radius of a boson star does not depend on its mass, 
this value should also fit  the radii of all
supermassive BHs. 
However, it obviously does not fit the observed
radius limit of   Sgr A$^*$,  $R<60$ AU, 
and even less it fits the radius of  a typical AGN
of 7 AU (section \ref{agn-mass}).

Hence, boson stars with a quartic self-interaction are ruled out as a unique explanation for 
compact supermassive objects at galactic centers and AGN.
Again, it is not excluded that some of the galactic centers
harbor a boson star.

\subsection{Dark-Energy Stars}
 Dark energy (DE) is a substance of negative pressure needed for 
 an accelerated cosmology.
 In order to achieve acceleration, DE must satisfy
 an equation of state $p_{\rm de}= w \rho_{\rm de}$, where $w$ may depend on
 the cosmological scale $a$ but at present ($a=1$) $w< -1/3$.
 In addition, one also expects that DE  satisfies the dominant energy condition, 
 in which case the equation of state satisfies
 $w\geq -1$. Nevertheless, matter with $w<-1$ dubbed ``phantom energy"
 \cite{cal2,cal} has recently received attention
 as it might  fit the most recent SN 1a data 
 slightly better than the  
 usual DE.
 
  In cosmological context,  $\rho_{\rm de}$ is homogeneous
 and it is normally assumed that DE does not cluster.
 However, it is not excluded that owing to gravitational
instability, small inhomogeneities 
(analogous to dark-matter inhomogeneities)  grow and build structure, e.g.,
 in the form  of spherical configurations. 
 Hence, we define {\em dark-energy stars} as 
 spherical solutions to Einstein's equation with matter
 described by the DE equation of state \cite{chap,lob3}.
It is worth noting that DE could be described by a scalar field
theory (quintessence)  with a suitably
chosen interaction potential and/or with a noncanonical kinetic energy term
in the Lagrangian. This scalar field theory would correspond to an effective
equation of state with desirable properties \cite{bil}.
In this way,  DE stars could be regarded as boson stars but with a rather
unusual self-interaction. 

 Here, we briefly  discuss two examples of DE stars: de Sitter gravastars and 
Chaplygin gravastars.
 
 \subsubsection{De Sitter  Gravastars}
 The simplest example of DE star is a gravitating 
 vacuum star or a {\em gravastar}.
 Chapline et al.\ \cite{chap1,chap2}  put forth an
interesting proposal based on analogies to condensed matter
systems where the effective general relativity was an emergent phenomenon.
Specifically, assuming the Schwarzschild exterior, they suggested that the sphere
 where the lapse function $\xi=g_{00}^{1/2}$ vanished marked a quantum phase
transition, $\xi$ increasing again at $r < 2GM$.
As this required negative pressure, the authors of \cite{chap2}
assumed the interior vacuum condensate to be described by de Sitter space
with the equation of state $p=-\rho$.

Subsequently, the idea of gravitational vacuum condensate, or `gravastar',
 was taken up by Mazur and Mottola \cite{mazu3,mazu4}
 and Dymnikova \cite{dym},
replacing the horizon with a shell of stiff matter astride
the surface at $r=2GM$.
Visser and Wiltshire \cite{viss4} and recently Carter \cite{car} also examined the stability
of the gravastar using the Israel thin-shell formalism \cite{isra5}.

Given the mass M, the interior is described by a solution,
similar to the the Einstein-de Sitter universe, with constant
density $\rho=\rho_0$
up to  the de Sitter radius
  $R_{\rm dS}=2M$ (actually slightly further out).
Hence, the interior is de Sitter, $\rho_0$ being the vacuum
energy density.
The lapse function is given by
\begin{equation}
\xi=\left(1-\frac{r^2}{R_{\rm dS}^2}\right)^{1/2} ,
\end{equation}
with
\begin{equation}
    R_{\rm dS} = \sqrt{\frac{3}{8\pi}} \rho_0^{-1/2}.
\end{equation}
 
In order to 
join
the interior solutions to a
Schwarzschild exterior at a spherical boundary of radius $R$, 
 it is necessary to put a thin spherical shell
at the boundary with a surface density 
and a surface tension
satisfying Israel's junction conditions \cite{isra5}.
As the pressure
does not vanish
at the boundary,
it must be  compensated by a negative surface tension of the shell. 

The gravastar has no horizon or singularity! Its surface is located 
at a radius just slightly
larger than the Schwarzschild radius $R=2M+\epsilon$, $\epsilon\sim 2l_{\rm Pl}$.
Hence, there is practically no observational way to distinguish a Schwarzschild BH from
a gravastar. Basically, any BH-like object observed in  nature may be a gravastar.

In spite of these attractive features the gravastar has been met with a cool reception.
Certainly, the assumption of a de Sitter interior presents a quandary:
on the one hand, the quantum phase transition would suggest that
the associated cosmological constant is a fundamental parameter;
on the other hand, to accommodate the mass range of supermassive black-hole
candidates, it must vary over some six orders of magnitude.
In addition, the notorious cosmological constant problem
is reversed for gravastars:
 why is the vacuum energy in the interior of a
gravastar so much larger than the observed vacuum energy density
in the universe? 
If we identify the most massive black-hole candidate
observed at
the center of M87, with mass
$M_{\rm max}=
3\times 10^9 M_{\odot}$, with the de Sitter gravastar,
then $\rho_0=(9.7 {\rm keV})^4$, to be contrasted with
the $(10^{-3} {\rm eV})^4$ values wanted for  cosmology.
Another question is
how does a gravastar form? The entropy of a stellar mass gravastar
is much less than the entropy of an ordinary star and this would require 
an extremely efficient cooling mechanism before  gravastars could form 
during stellar collapse
\cite{abr}.

\subsubsection{Chaplygin Gravastars}

Consider a particular form of DE with a rather peculiar 
equation of state 
\begin{equation}
p = - \; \frac{A}{\rho}.
\label{eq941}
\end{equation}
Equation (\ref{eq941}) describes the Chaplygin gas which,
for $\rho \geq \sqrt{A}$, has attracted some attention as a
dark-energy candidate \cite{bil,kamen7}.
Consider a  self-gravitating Chaplygin gas 
and look for static solutions. 
In particular, we look for static Chaplygin gas configurations in the phantom 
($w<-1$) regime,
 i.e., when 
\begin{equation}
\rho < \sqrt{A} \, .
\label{eq942}
\end{equation}
We  show that these configurations could
provide an alternative scenario for compact massive objects at galactic
centers \cite{bil2}. Moreover, equation (\ref{eq941})
yields the de Sitter gravastar solution
in the limit when the central density of the static solution
approaches the value $\sqrt{A}$.

Combining  (\ref{eq941}) with the TOV equations
(\ref{eq48}),
and (\ref{eq49}),
 one has
\begin{equation}
\rho' =  \left( 1 - \frac{\rho^{2}}{A} \right)
\left( \frac{ \rho {\cal{M}} - 4 \pi A r^{3} }{r (r - 2 {\cal{M}} ) }  \right) \, ,
\label{eq943}
\end{equation}
\begin{equation}
\frac{d{\cal{M}}}{dr}=4\pi r^2 \rho.
\end{equation}

\begin{figure}
\begin{center}
\includegraphics[width=.5\textwidth,trim= 0 3cm 0 4cm]{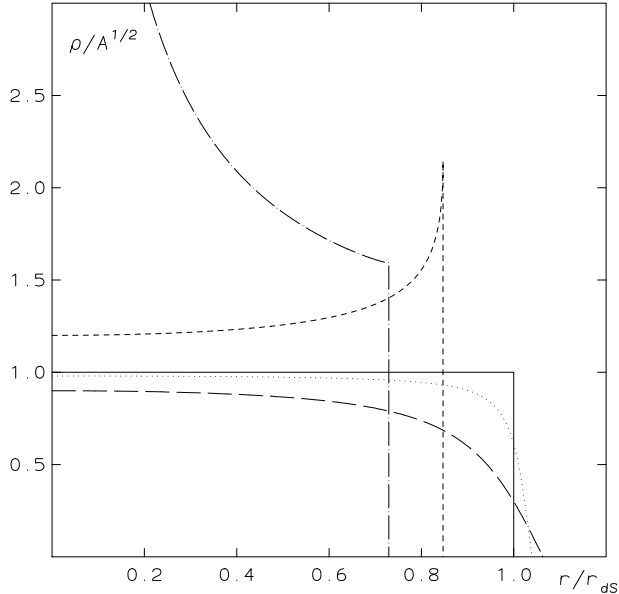}
\caption{
 Density profile of the Chaplygin gravastar for
 $\rho_0/\sqrt{A}= 1.2$ (short dashed), 0.98 (dotted),
 0.9 (long dashed).
The limiting singular solution with the 
 $r^{-2/3}$ behavior at small $r$ is represented by the dot-dashed
 and the de Sitter gravastar by the solid line.
}
\label{fig2}
\end{center}
\end{figure}

In Fig.\ \ref{fig2}  we exhibit
the resulting
$\rho (r)$ for
selected values of
$\rho_{0} / \sqrt{A}$.
The solutions depend essentially
on the magnitude of $\rho_0$ relative to
 $\sqrt{A}$. In the following we summarize the properties of three classes of
 solutions corresponding to whether $\rho_0$ is larger, smaller, or
 equal to $\sqrt{A}$.

{\bf i)} For $\rho_0>\sqrt{A}$, the density
$\rho$ increases and the lapse function $\xi$ decreases with $r$
starting  from the origin
up to the black-hole horizon radius
 $R_{\rm bh}$,
    where $2G{\cal{M}}(R_{\rm bh})=R_{\rm bh}$.
In the limit
$\rho_0 \rightarrow \infty$,
a limiting
solution exists with a singular behavior
\begin{equation}
    \rho(r) \simeq (\frac{7 A}{18\pi G r^2})^{1/3}
\label{eq944}
\end{equation}
  near the origin.

{\bf ii)} For $\rho_0<\sqrt{A}$,
both
  $\rho$ and $\xi$ decrease with $r$ up to the radius $R_0$ where they vanish.
  At that point the pressure $p$ blows up to $-\infty$ 
     owing to (\ref{eq941}).
    The enclosed mass
${\cal M}$ is always less than $r/(2)$,  never reaching the black-hole horizon,
i.e., the radius where $2{\cal M}(r)=r$.

{\bf iii)} For $\rho_0=\sqrt{A}$,
the density $\rho$ remains
 constant equal to $\sqrt{A}$ up to  the de Sitter radius
  $R_{\rm dS}=2M$.
Hence, we recover the de Sitter solution precisely as in the gravastar case.

As $\rho_0 \rightarrow \sqrt{A} $  from above or from below,
solutions {\bf i)} or
{\bf ii)}, respectively, converge   to
{\bf iii)}
except at the endpoint.
The lapse function in {\bf iii)}
 joins the Schwarzschild solution outside
continuously,
whereas in {\bf i)} and {\bf ii)} it happens discontinuously.

As in the case of a de Sitter gravastar, in order to 
join
our interior solutions to a
Schwarzschild exterior at a spherical boundary of radius $R$, 
 it is necessary to put a thin spherical shell\footnote{It has  recently
 been demonstrated that the joining can be made continuous
without the presence of a thin shell 
for a gravastar made of the fluid with an 
anisotropic pressure \cite{cat,ben}.}
at the boundary with a surface density 
and a surface tension
satisfying Israel's junction conditions \cite{isra5}.

Case {\bf ii)}, together with {\bf iii)}, is of particular
interest as we would like to interpret the supermassive
compact 
dark
objects at the galactic centers in terms of
phantom energy rather than in terms of a classical black hole.
It is natural to
assume that the most massive such object is described by
the de Sitter gravastar, i.e., solution {\bf iii)}
(depicted by the solid line in Fig.\ \ref{fig2}).
The radius of this object is $R_{\rm dS}$ equal to the Schwarzschild radius
 $2M_{\rm max}$.
Clearly,  solutions belonging to class {\bf ii)}, can fit
all masses $M<M_{\rm max}$.
However, for the phenomenology  of supermassive
galactic centers
it is important to find, at least approximately,
the mass-radius relationship for these solutions.
This may be done in the low central density approximation, i.e.,
$\rho_0 \ll \sqrt{A}$,
which is similar to the Newtonian approximation but,
in contrast to the  Newtonian approximation, one cannot neglect the pressure  term
in  (\ref{eq48}). Moreover,
 as may be easily shown, in this approximation
${\cal{M}} \ll r^3 p$, so that the pressure term becomes dominant.
Next, neglecting $2 \cal M$ with respect to $r$, as in the usual
Newtonian approximation,
 equation (\ref{eq943})
simplifies to
$
\rho' =
 4 \pi A r $,
with the solution
\begin{equation}
\rho = \rho_0 \left(
1-\frac{r^2}{R_0^2}\right) \, ;
\;\;\;\;\;
R_0^2= \frac{\rho_0}{2 \pi  A} \, ,
\end{equation}
which gives a mass-radius relation
\begin{equation}
\frac{M}{R_0^5}= \frac{16\pi^2}{15}A={\rm constant}.
\label{eq945}
\end{equation}
The mass-radius relationship
$M \propto R_0^{5}$ which phantom gravastars obey, offers the prospect
of unifying the description of all supermassive compact dark objects 
at the galactic centers, as Chaplygin-gas phantom gravastars
with masses ranging from
$M_{\rm min} = 10^{6}  M_{\odot}$ to
$M_{\rm max} = 3 \times 10^{9}  M_{\odot}$.
Indeed, assuming that the most massive compact dark object, observed at
the center of M87, is a Chaplygin  phantom gravastar near the 
de Sitter gravastar limit, with
$R_{\rm max} = 2M_{\rm max} = 8.86 \times 10^{9} \mbox{km} = 8.21 \; \mbox{lhr}$,
the compact dark object at the center of our Galaxy, with mass
$M_{\rm GC} =  3 \times 10^{6} M_{\odot}$, would have a radius $R_{\rm GC}$ = 2.06 lhr
if the scaling law (\ref{eq945}) holds. This radius is well below the distances
of the closest approach to Sgr A$^{*}$ which the stars SO-2
($R_{\rm min}$ = 17 lhr = 123 AU, \cite{schod1,eis}) and 
SO-16 ($R_{\rm min}$ = 8.32 lhr = 60 AU \cite{ghez3,ghez4})
recently had and
beyond which the Keplerian nature of the gravitational potential of Sgr A$^{*}$
is well established.

\appendix
\section{Basic General Relativity}
\label{relativity}
\subsection{Geometry}
 The 
geometry of spacetime is described by the metric
\begin{equation}
ds^2=g_{\mu\nu}dx^\mu dx^\nu \, ,
\label{eq202}
\end{equation}
where  $g_{\mu\nu}$ is the metric tensor.
The spacetime curvature is defined through
the Riemann curvature tensor 
\begin{equation}
R_{\tau\sigma\beta\gamma}=g_{\tau\alpha}R^{\alpha}_{\; \sigma\beta\gamma} \: ,
\end{equation}
 defined by
\begin{equation}
R^{\alpha}_{\;
\delta\beta\gamma}=\partial_{\gamma}\Gamma^{\alpha}_{\beta\delta} -
\partial_{\beta}\Gamma^{\alpha}_{\delta\gamma} 
+ \Gamma^{\alpha}_{\rho\gamma}\Gamma^{\rho}_{\beta\delta}-
\Gamma^{\alpha}_{\rho\beta}\Gamma^{\rho}_{\gamma\delta} \: , 
\end{equation}  
where  
\begin{equation}
\Gamma^{\sigma}_{\mu\nu}=\frac{1}{2}g^{\lambda\sigma}[
\partial_{\mu}g_{\nu\lambda}+ \partial_{\nu}g_{\mu\lambda} -
\partial_{\lambda}g_{\mu\nu} ]
\end{equation}
are the Christoffel symbols. 
A contraction of two indices gives  
 the {\em Ricci tensor} 
\begin{equation}
R_{\alpha\beta}=g^{\lambda\nu}R_{\lambda\alpha\nu\beta} 
\end{equation}
and a further contraction of the
 Ricci tensor gives 
 the {\em scalar curvature} 
\begin{equation}
{\cal R}=g^{\alpha\beta} R_{\alpha\beta} \: . 
\end{equation}

\subsection{Covariant Derivative}

A covariant derivative $\nabla_\mu$ of a field $Y$
(in general, $Y$ may be a tensor of any rank) is defined by
\begin{equation}
DY = (\nabla_\mu Y)dx^{\mu}\equiv Y_{;\mu}dx^{\mu}\: ,
\end{equation} 
where $DY$ is an infinitesimal difference between the value
of the field $Y$ at the point $x^\mu+dx^\mu$ and the quantity
$Y(x^\mu)$ parallelly displaced  from $x^\mu$ to
$x^\mu+dx^\mu$. Hence, $DY$ consists of two parts:
one is the change of $Y$ due to the parallel displacement
and the other is the  difference
$dY=Y(x^\mu+dx^\mu)-Y(x^\mu)$ due to the functional dependence on $x^\mu$.
The latter part is basically related to the ordinary partial differentiation.
 The difference due to the parallel displacement
is related to the curvature of  spacetime and  
depends on the tensor nature of $Y$.
For a scalar, vector, and second-rank tensor one finds:
\cite{lan2}
\begin{itemize}
\item
scalar $\varphi$
\begin{equation}
\varphi_{;\mu}=\varphi_{,\mu}\: ,
\end{equation} 
\item 
vector $V_\mu$
\begin{equation}
V_{\mu;\nu}=V_{\mu,\nu}-\Gamma^\rho_{\mu\nu}V_\rho\, ,
\;\;\;\;
{V^\mu}_{;\nu}={V^\mu}_{,\nu}+\Gamma^\mu_{\rho\nu} V^\rho\, .
\end{equation} 
\item
tensor $A_{\mu\nu}$
\begin{equation}
A_{\mu\nu;\rho}=A_{\mu\nu,\rho}-\Gamma^\sigma_{\mu\rho}A_{\sigma\nu}-
\Gamma^\sigma_{\nu\rho}A_{\mu\sigma}\, ,
\;\;\;\;
{A^{\mu\nu}}_{;\rho}={A^{\mu\nu}}_{,\rho}+\Gamma^\mu_{\sigma\rho}A^{\sigma\nu}
+\Gamma^\nu_{\sigma\rho}A^{\mu\sigma} \, .
\end{equation} 
\end{itemize}
The covariant d'Alembertian is given by
\begin{equation}
\mbox{\Large $\Box$}\equiv g^{\mu\nu}\nabla_\mu\nabla_\nu=
 \frac{1}{\sqrt{-\det g}}\,
 \partial_\mu  \left (\sqrt{-\det g }\, g^{\mu \nu }
 \partial_\nu \right ).
\end{equation}

Here we have used the usual convention in which a subscript $,\mu$
denotes an ordinary partial derivative
and $;\mu$ denotes the covariant derivative.
\subsection{Geodesics}
\label{geodesics}
Geodesics are the ``shortest possible lines'' one can 
draw in curved geometry. Given a covariant derivative operator $\nabla_\mu$,
we define a {\em geodesic} to be a curve whose tangent vector is parallel
propagated along itself, i.e., a curve whose tangent $u^{\mu}$ satisfies the equation
\begin{equation}
u^{\nu}{u^\mu}_{;\nu}=0\, .
\label{eq204}
\end{equation} 
A massive particle moves along a timelike geodesic, $u_\mu$ being its 
velocity. A massless particle moves along a null geodesic in which case
the vector $u_\mu$ is null. 

\subsection{Isometries and Killing Vectors}
\label{isometries}
A vector field $k^\mu$ that generates one parameter group of isometries, i.e.,
 one parameter  group of transformations that leave the metric invariant, 
 is called a {\em Killing vector}. 
A Killing vector $k^{\mu}$  satisfies the Killing equation
\begin{equation}
k_{(\mu;\nu)}\equiv k_{\mu;\nu}+k_{\nu;\mu}=0.
\label{eq805}
\end{equation} 
It is convenient to represent the Killing vector $k^\mu$ as a
differential operator $k$,
\begin{equation}
k=k^\mu\partial_\mu .
\end{equation} 
For a vector field  $k$, local coordinates can be found such
that
\begin{equation}
k=\frac{\partial}{\partial x} \, ,
\end{equation}
where $x$ is one of these coordinates, e.g., $x^4\equiv x$. 
In such a coordinate system, $k^\mu=\delta^\mu_4$ and
\begin{equation}
k_{(\mu;\nu)}=
\frac{\partial g_{\mu\nu}}{\partial x}=0.
\end{equation} 
Hence, one can say that $k_{\mu}$ is Killing if $g_{\mu\nu}$ is 
independent of $x$.
\vskip \baselineskip
\noindent
{\bf Examples}
\vskip \baselineskip
Consider the spherical coordinate system
$(t,r,\theta , \phi)$.
Suppose the metric components are independent of $t$ and $\phi$. 
Then there exist
 two Killing vectors:
 \begin{itemize}
 \item
 the generator of time translations $\xi=\partial/\partial t$
 \begin{equation}
\xi^{\mu}= \delta^\mu_0; \;\;\;\;\; 
\xi_{\mu}= g_{\mu 0} \, ,
\label{eq800}
\end{equation} 
 \item
 the generator of axial isometries $\psi=\partial/\partial \phi$.
 In spherical coordinates 
 \begin{equation}
\psi^{\mu}= \delta^\mu_\phi; \;\;\;\;\; 
\psi_{\mu}= g_{\mu \phi} \, .
\label{801}
\end{equation} 
\end{itemize}
\subsection{Constants of Motion}
\label{constants}
The following proposition relates  Killing vectors 
to constants of motion.
\begin{proposition}
 Let $\chi^\mu$ be a Killing vector field
  and let $\gamma$ be a geodesic
 with tangent 
   $u^\mu$.
 Then the quantity
$\chi_{\mu}u^{\mu}$
 is constant along $\gamma$.
 \label{killing}
\end{proposition}
{\bf Proof}:
We have
\begin{equation}
u^\nu(\chi_{\mu}u^{\mu})_{;\nu}= 
u^{\mu}u^{\nu}\chi_{\mu ;\nu}+ \chi_\mu u^\nu {u^\mu}_{;\nu} =0.
\end{equation}
The first term  
 vanishes by the Killing equation (\ref{eq805}) and the second term
vanishes by the geodesic equation (\ref{eq204}) . Hence
\begin{equation}
 u^{\nu}(\chi^{\mu}u_{\mu})_{;\nu}
 =0,
\end{equation}
as was to be proved. {\Large $\Box$}
\vskip \baselineskip
\noindent
{\bf Examples}
\vskip \baselineskip
Consider a stationary  axially symmetric space, as in  the example 
discussed in appendix \ref{isometries} with two Killing vectors, 
$\xi=\partial/\partial t$ and $\psi=\partial/\partial\phi$,
corresponding to time translation and axial isometries, re\-spec\-tive\-ly.
 In addition, assume the asymptotic flatness, i.e.,
$g_{\mu\nu}\rightarrow$ diag$(1,-1,-r^2, -r^2 \sin^2 \theta)$
as $r\rightarrow \infty$.
For a particle of mass $m$ moving along a timelike geodesic, the two
conserved quantities along the geodesics are
 \begin{itemize}
 \item
 the energy  
 \begin{equation}
E= m u^{\mu}\xi_{\mu}= \xi_{\mu}q^{\mu}= g_{0\mu} q^\mu= 
\left. q^0\right|_{\infty}\, ,
\label{eq808}
\end{equation} 
 \item
 the z-component of the angular momentum  
 \begin{equation}
L=-m u^{\mu}\psi_{\mu}= -\psi_{\mu}q^{\mu}=-g_{\phi\mu} q^\mu= \left. 
r^2\sin^2\theta q^\phi\right|_{\infty}\, .
\label{eq809}
\end{equation} 

\end{itemize}

\subsection{Einstein's  Equations}
General relativity relates the geometry of spacetime
to matter through Einstein's field equations
 \begin{equation}
R_{\mu\nu}-\frac{1}{2}g_{\mu\nu} {\cal R}=-8\pi T_{\mu\nu} \, ,
\label{eq201}
\end{equation}
where $T_{\mu\nu}$ is the {\em energy-momentum tensor}.

\section{Basic Fluid Dynamics}
\label{fluid}
Consider
 a perfect gravitating relativistic fluid.
We denote by
 $u_{\mu}$ ,
$p$, $\rho$, $n$,  and $\sigma$ the velocity, pressure,
energy density, particle number density, and
entropy density of the fluid.
The energy-momentum tensor of a perfect fluid is
given by
\begin{equation}
T_{\mu\nu}=(p+\rho) u_{\mu}u_{\nu}-p g_{\mu\nu}   ,
\label{eq001}
\end{equation}
where
 $g_{\mu\nu}$ is the metric tensor
with the Lorentzian signature
$(+---)$. Hence, in this convention, we have
\begin{equation}
 u^{\mu}u_{\mu}=  g_{\mu\nu}
 u^{\mu}u^{\nu}=1.
\label{eq101}
\end{equation}

The particle number conservation is described by 
the continuity equation
\begin{equation}
(nu^{\mu})_{;\mu}=
\frac{1}{\sqrt{-g}}
\partial_{\mu}({\sqrt{-g}}\,  n u^{\mu})=0 .
\label{eq002}
\end{equation}
The energy-momentum
conservation
\begin{equation}
{T^{\mu\nu}}_{;\nu}=0
\label{eq102}
\end{equation}
  applied to  (\ref{eq001}) yields
 the relativistic generalization of Euler's equation \cite{lan}
\begin{equation}
(p+\rho)u^{\nu}u_{\mu;\nu}
-\partial_{\mu}p
+u_{\mu}u^{\nu}
\partial_{\nu}p  =0.
\label{eq003}
\end{equation}
%

\subsection{Fluid Velocity}
It is convenient to
  parameterize the four-velocity of the fluid
   in terms of
  three-velocity components.
  To do this, we
   use the projection operator
$g_{\mu\nu}-
t_{\mu}t_{\nu}$,
which projects a vector into
the subspace orthogonal to
the time-translation Killing vector
 $\xi^{\mu}=(1;\vec{0})$ ,
 where $t_{\mu}$ is the unit vector
\begin{equation}
t^{\mu}=
\frac{\xi^{\mu}}{\sqrt{\xi^{\nu}\xi_{\nu}}}=
\frac{
\delta^{\mu}_{0}
}{\sqrt{g_{00}}}\,  ;
 \;\;\;\;\;\;\;
t_{\mu}=
\frac{\xi_{\mu}}{\sqrt{\xi^{\nu}\xi_{\nu}}}=
\frac{
g_{\mu 0}
}{\sqrt{g_{00}}}\, .
\label{eq130}
\end{equation}
We split up the vector $u_{\mu}$ in  two parts: one parallel
 with and the other orthogonal to
 $t_{\mu}$:
\begin{equation}
u_{\mu}=
\gamma t_{\mu}+
(g_{\mu\nu}-
t_{\mu}t_{\nu})
u^{\nu}  ,
\label{eq230}
\end{equation}
where
\begin{equation}
\gamma=
 t^{\mu}
 u_{\mu} \, .
\label{eq330}
\end{equation}

From (\ref{eq230})
with (\ref{eq130}) we find
(see, e.g., appendix of \cite{bil5})
\begin{eqnarray}
\nonumber \\
u^{\mu}
 \!&\!=\!&\!
\gamma \left(
 \frac{1}{\sqrt{g_{00}}}
-\frac{g_{0j}v^j}{g_{00}};v^i
\right)   ,      \nonumber \\
u_{\mu}
 \!&\!=\!&\!
\gamma \left(
 \sqrt{g_{00}};
\frac{g_{0i}}{\sqrt{g_{00}}}-v_i
\right) ,
\label{eq031}
\end{eqnarray}
where 
\begin{equation}
v_i=\gamma_{ij} v^j, \;\;\;\;\; v^2=v^i v_i ,
\;\;\;\;\;\;
\gamma^2= (1-v^2)^{-1},
\label{eq032}
\end{equation}
with the induced
three-dimensional spatial metric:
\begin{equation}
\gamma_{ij}=
\frac{g_{0i}g_{0j}}{g_{00}}-g_{ij}   \, ;
 \;\;\;\;\;
 i,j=1,2,3.
\label{eq033}
\end{equation}
Since $u^{\mu}$ and $t^{\mu}$ are timelike unit vectors,
 a consequence of
 (\ref{eq230}) is that
 $\gamma \geq 1$ and hence $0 \leq v^2 < 1$.
%

\subsection{Hydrostatic Equilibrium}
From Euler's equation (\ref{eq003}) we can derive the condition of 
hydrostatic
equilibrium. We can 
use the comoving frame of reference in which the fluid 
velocity takes the form
\begin{equation}
u^{\mu}=
\frac{
\delta^{\mu}_0
}{\sqrt{g_{00}}}\,  ;
 \;\;\;\;\;\;\;
u_{\mu}=
\frac{
g_{\mu 0}
}{\sqrt{g_{00}}}\, .
\label{eq024}
\end{equation}
In equilibrium the  metric is static; all components are independent of time,
and the mixed components $g_{0i}$ are zero. Equation
(\ref{eq003}) then gives
\begin{equation}
(\rho+p) \Gamma_{\mu0}^0 u^0u_0=(\rho+p) \frac{1}{2}
g^{00}\partial_\mu g_{00}
=-\partial_\mu p
\label{eq025}
\end{equation}
or
\begin{equation}
\partial_{\mu}p=-(p+\rho)g_{00}^{-1/2}\partial_{\mu}g_{00}^{1/2} .
\label{eq17}
\end{equation}

\section{Basic Thermodynamics}
\label{thermo}
Consider a nonrotating fluid consisting of $N$ particles
in equilibrium
at nonzero temperature.
A canonical ensemble is subject to the constraint
that the number of particles
\begin{equation}
\int_{\Sigma} n\, u^{\mu}d\Sigma_{\mu}
=N
\label{eq26}
\end{equation}
should be fixed.
The spacelike hypersurface
$\Sigma$ that contains
the fluid is orthogonal to the time-translation
Killing vector field $\xi^{\mu}$.
In equilibrium $\xi^{\mu}$ is related to the velocity of the fluid.
\begin{equation}
\xi^{\mu}=\xi u^{\mu}\, ;  \;\;\;\;\;\;
\xi=(\xi^{\mu}\xi_{\mu})^{1/2}.
\label{eq50}
\end{equation}

It may be shown that those and only those configurations will be in equilibrium 
for which
the free energy 
assumes a minimum \cite{bilGRG}. 
The canonical free energy is defined
as \cite{bilGRG,gib}.
\begin{equation}
F=M-\int_{\Sigma} T\sigma \, \xi^{\mu}d\Sigma_{\mu} \, ,
\label{eq30}
\end{equation}
where $M$ is the total mass as measured from infinity.
The entropy density is obtained using the standard thermodynamic relation
\begin{equation}
\sigma=\frac{1}{T}(p+\rho-\mu n).
\label{eq16}
\end{equation}

\subsection{Tolman Equations}
\label{tolman}
The temperature $T$ and the chemical potential $\mu$ are
  metric-dependent local quantities.
  Their spacetime dependence may be derived from
the equation of hydrostatic equilibrium (\ref{eq17})
\cite{tol,lan}
and the thermodynamic identity (Gibbs-Duhem relation)
\begin{equation}
 d\frac{p}{T}=
 n d\frac{\mu}{T}-\rho d\frac{1}{T}.
\label{eq18}
\end{equation}
The crucial condition is that the heat flow and diffusion should vanish
\cite{isr}
\begin{equation}
  \frac{\mu}{T}={\rm const} ,
\label{eq19}
\end{equation}
which may be  derived from the 
physical requirement that the rate of
entropy change with
particle number at fixed energy density
should be  constant, i.e.,
\begin{equation}
\left. \frac{\partial \sigma}{\partial n}\right|_{\rho}={\rm const} ,
\label{eq32}
\end{equation}
where `const' is independent of $\rho$.
From (\ref{eq16}) and (\ref{eq18}) we obtain
\begin{equation}
d\sigma=  \frac{1}{T}d\rho-\frac{\mu}{T}dn
\label{eq29}
\end{equation}
and hence
\begin{equation}
\left.\frac{\partial \sigma}{\partial n}\right|_{\rho}=-\frac{\mu}{T}.
\label{eq31}
\end{equation}
from which (\ref{eq19}) immediately follows.
Next, equation (\ref{eq19}), 
together with (\ref{eq17}) and (\ref{eq18}), implies
the well-known Tolman equations
\begin{equation}
T g_{00}^{1/2}=T_0\, ; \;\;\;\;\;\;
\mu g_{00}^{1/2}=\mu_0  \, ,
\label{eq21}
\end{equation}
where $T_0$ and $\mu_0$ are constants equal,
respectively, to the
temperature and the chemical potential at infinity.
In a grand-canonical ensemble,
$T_0$ and $\mu_0$ may be chosen arbitrarily.
In a canonical ensemble,  $\mu_0$ is
an implicit functional of $\xi$ because of
the constraint (\ref{eq26}) that the total number of particles
should be fixed.
\subsection{Fermi Distribution}
Consider a
gas consisting of $N$ fermions with the mass
$m$ contained within  a hypersurface $\Sigma$.
The equation of state may be represented
in a parametric form using
the well-known momentum integrals over
 the Fermi distribution function
\cite{ehl}
\begin{equation}
n   = g \int^{\infty}_{0} \frac{d^3q}{(2\pi)^3}\,
\frac{1}{1+e^{E/T-\mu/T}} \, ,
\label{eq13}
\end{equation}
\begin{equation}
\rho = g \int^{\infty}_{0} \frac{d^3q}{(2\pi)^3}\,
\frac{E}{1+e^{E/T-\mu/T}} \, ,
\label{eq14}
\end{equation}
\begin{equation}
p = g T \int^{\infty}_{0} \frac{d^3q}{(2\pi)^3}\,
\ln (1+e^{-E/T+\mu/T}) \, ,
\label{eq15}
\end{equation}
where
 $E=\sqrt{m^2+q^2}$ and
$T$ and $\mu$ are the local temperature and the local chemical potential,
respectively,
defined by Tolman's equations 
 (\ref{eq21}).
By partial integration, the last equation may be written as
\begin{equation}
p = g \int^{\infty}_{0} \frac{d^3q}{(2\pi)^3}\,
\frac{q^2}{3E}\frac{1}{1+e^{E/T-\mu/T}} \, ,
\label{eq314}
\end{equation}
The integer $g$ denotes the spin degeneracy factor. 
In most applications we
take $g=2$
(spin up and spin down).
Strictly speaking, in each equation (\ref{eq13})-(\ref{eq314})
one should also add
the antiparticle term which is of the same form
as the corresponding right-hand side of (\ref{eq13})-(\ref{eq314})
with $\mu$ replaced by $-\mu$. 
However, the contribution of antiparticles in  astrophysical
objects 
is almost always negligible\footnote{One exception is 
 neutrino stars made of Dirac type neutrinos. 
There, the numbers of neutrinos and antineutrinos are equal
and separately conserved, hence g=4}.

 Equations (\ref{eq16}) and (\ref{eq29}) may be combined to 
yield another
useful thermodynamic identity
\begin{equation}
 dw=
 T  d(\frac{\sigma}{n})+\frac{1}{n}dp,
\label{eq004}
\end{equation}
with
$w=(p+\rho)/n$ being the specific enthalpy.
\subsection{Isentropic Fluid}
Euler's equation is simplified if one
restricts consideration to
an isentropic flow.
A flow is said to be {\em isentropic} when the specific entropy
$\sigma/n$ is constant, i.e.,when
\begin{equation}
\partial_{\mu}(\frac{\sigma}{n})=0 .
\label{eq103}
\end{equation}

A flow may in general have a nonvanishing
vorticity
$\omega_{\mu\nu}$
defined as
\begin{equation}
\omega_{\mu\nu}= h^{\rho}_{\mu}
h^{\sigma}_{\nu} u_{[\rho;\sigma]},
\label{eq203}
\end{equation}
where
\begin{equation}
h^{\mu}_{\nu} =
\delta^{\mu}_{\nu}-
u^{\mu}u_{\nu}
\label{eq303}
\end{equation}
is the projection operator
which projects an arbitrary vector in spacetime
into its component in the subspace orthogonal to
$u^{\mu}$.
A flow with vanishing vorticity, i.e., when
\begin{equation}
\omega_{\mu\nu}= 0,
\label{eq403}
\end{equation}
 is said to be {\em irrotational}.
 In the following we assume that the flow is
 isentropic and irrotational.

As a consequence of
equation (\ref{eq103}) and the
thermodynamic identity (\ref{eq004}),
equation (\ref{eq003})
simplifies to
\begin{equation}
 u^{\nu}(wu_{\mu})_{;\nu}
-\partial_{\mu}w=0.
\label{eq005}
\end{equation}
Furthermore, for an isentropic irrotational flow,
 equation (\ref{eq403}) implies \cite{tau78}
\begin{equation}
 (wu_{\mu})_{;\nu}
 -(wu_{\nu})_{;\mu}=0.
\label{eq006}
\end{equation}
In this case, we may introduce a scalar function
$\varphi$ such that
\begin{equation}
  wu_{\mu}= -\partial_{\mu} \varphi ,
\label{eq007}
\end{equation}
where the minus sign is chosen for convenience.
Obviously,
the quantity $wu^{\mu}$
in the form (\ref{eq007})
satisfies equation (\ref{eq006}).
Solutions of this form are the relativistic analogue
of potential flow in nonrelativistic fluid dynamics
\cite{lan}.
 \subsection{Degenerate Fermi Gas}
 \label{degenerate}
In the limit $T\rightarrow 0$,
 the Fermi distribution in
(\ref{eq13})-(\ref{eq15})
becomes a step function that  yields
an elementary integral with the upper limit
equal to the Fermi
momentum
\begin{equation}
q_{\rm F}=\sqrt{\mu^2 -m^2}=mX .
\label{eq95}
\end{equation}
The equation of state can be expressed
in terms of elementary functions
 of $X$.
With $g=2$, we find
\begin{equation}
n=2\int_0^{q_{\rm F}} \frac{d^3q}{(2\pi)^3}=\frac{1}{3\pi^2}m^3X^3,
\label{eq96}
\end{equation}
\begin{equation}
\rho=2\int_0^{q_{\rm F}} E\frac{d^3q}{(2\pi)^3}=\frac{1}{8\pi^2}m^4 \left[X(2X^2+1)\sqrt{X^2+1}
-{\rm Arsh}X\right] ,
\label{eq97}
\end{equation}
\begin{equation}
p=2\int_0^{q_{\rm F}} \frac{q^2}{3E}\frac{d^3q}{(2\pi)^3}=\frac{1}{8\pi^2}m^4 \left[X(\frac{2}{3}X^2-1)\sqrt{X^2+1}
+{\rm Arsh}X\right] .
\label{eq98}
\end{equation}

There are two important limits:
\begin{itemize}
\item
Nonrelativistic limit, $X\ll 1$
\begin{equation}
\rho=\frac{1}{3\pi^2}m^4 \left(X^3+\frac{3}{10}X^5-\frac{3}{56}X^7 ...\right),
\label{eq197}
\end{equation}
\begin{equation}
p=\frac{1}{15\pi^2}m^4 \left(X^5-\frac{5}{14}X^7+\frac{5}{24}X^9 ...\right),
\label{eq198}
\end{equation}
Obviously, in this limit, $p\ll \rho$. 
\item
Ultrarelativistic limit
\begin{equation}
\rho=\frac{1}{4\pi^2}m^4 \left(X^4+X^2-\frac{1}{2}\ln 2X ...\right),
\label{eq297}
\end{equation}
\begin{equation}
p=\frac{1}{12\pi^2}m^4 \left(X^4-X^2+\frac{3}{2}\ln 2X ...\right).
\label{eq298}
\end{equation}
Retaining the dominant terms,  these equations yield
the  well-known equation of state for an ultrarelativistic gas
\begin{equation}
  p= \frac{\rho}{3} \, ,
\end{equation}  
which also holds for massless Bose and Fermi gases at finite temperature.
\end{itemize}
 \subsection{Polytropic Gases}
 \label{polytropic}
A gas of particles of mass $m$ is called {\em polytropic}
if the equation of state may be written in the form 
\begin{equation}
p={\cal K} n^{\Gamma},
\label{eq216}
\end{equation}
where ${\cal K}$ is a constant that depends on $m$.
Using this equation and
the thermodynamic identity (\ref{eq004})
the energy density $\rho$ and the entropy density $\sigma$ 
may also be expressed
in terms of $n$.
For an isentropic flow, it follows
\begin{equation}
\rho=m n+\frac{\cal K}{\Gamma-1} n^{\Gamma}.
\label{eq317}
\end{equation}
A degenerate Fermi gas  approaches a polytropic equation of state in 
both the nonrelativistic and the extreme-relativistic limits:
\begin{itemize}
\item
Nonrelativistic limit.  
 Retaining only the dominant term
in (\ref{eq198})
we find
\begin{equation}
p= \frac{3^{2/3}\pi^{4/3}}{5 m} n^{5/3} .
\label{eq398}
\end{equation}
\item
Ultrarelativistic limit. In this case, the dominant term in  (\ref{eq298})
yields
\begin{equation}
p= \frac{3^{1/3}\pi^{2/3}}{4}n^{4/3} .
\label{eq399}
\end{equation}
\end{itemize}

\subsection*{Acknowledgments}
I thank the organizers of the school for the invitation and hospitality. I am grateful to Andreas M\"uller for his valuable comments
which helped me improve the manuscript. This work  was supported by
 the Ministry of Science and Technology of the
 Republic of Croatia under Contract
 No. 0098002.

\end{document}